\documentclass[fleqn,usenatbib]{mnras}

\usepackage{newtxtext,newtxmath}
\usepackage[T1]{fontenc}

\DeclareRobustCommand{\VAN}[3]{#2}
\let\VANthebibliography\thebibliography
\def\thebibliography{\DeclareRobustCommand{\VAN}[3]{##3}\VANthebibliography}

\usepackage{graphicx}	% Including figure files
\usepackage{amsmath}	% Advanced maths commands
\usepackage[shortlabels]{enumitem}   % For enumerate label
\PassOptionsToPackage{hyphens}{url}
\usepackage{tikz,xcolor,hyperref} % For ORCID
\newcommand{\refbf}{}  %{\color{red}} % {\bf}  % {}

\setlist[enumerate]{
  labelsep=8pt,
  labelindent=0\parindent,
  itemindent=0pt,
  leftmargin=*,
  before=\setlength{\listparindent}{-\leftmargin},
}

\definecolor{lime}{HTML}{A6CE39}
\DeclareRobustCommand{\orcidicon}{%
    \begin{tikzpicture}
    \draw[lime, fill=lime] (0,0) 
    circle [radius=0.16] 
    node[white] {{\fontfamily{qag}\selectfont \tiny ID}};
    \draw[white, fill=white] (-0.0625,0.095) 
    circle [radius=0.007];
    \end{tikzpicture}
    \hspace{-2mm}
}

\newcommand{\orcidJJT}{\href{https://orcid.org/0000-0002-1860-0886}{\orcidicon}}
\newcommand{\orcidChrisW}{\href{https://orcid.org/0000-0002-4569-016X}{\orcidicon}}
\newcommand{\orcidJT}{\href{https://orcid.org/0000-0003-2858-9657}{\orcidicon}}

\newcommand{\orcidSamuel}{\href{https://orcid.org/0000-0001-9372-4611}{\orcidicon}}
\title[Timescales of Accretion Discs]{Timescales of Quasar Accretion Discs from Low to High Black Hole Mass and a Turnover at the High Mass End}

\author[Ch. Wolf et al.]
{Christian Wolf$^{1,2}$\orcidChrisW \thanks{E-mail: christian.wolf@anu.edu.au},
Samuel Lai$^{3}$\orcidSamuel,
Ji-Jia Tang$^{4,1}$\orcidJJT, %\thanks{E-mail: ji-jia.tang@anu.edu.au, tang.ji_jia.gf@ehime-u.ac.jp}
John Tonry$^5$\orcidJT \\
$^1$Research School of Astronomy and Astrophysics, Australian National University, Cotter Road, Weston Creek ACT 2611, Australia \\
$^2$Centre for Gravitational Astrophysics (CGA), Australian National University, Building 38 Science Road, Acton ACT 2601, Australia \\
$^3$Commonwealth Scientific and Industrial Research Organisation (CSIRO), Space \& Astronomy, P. O. Box 1130, Bentley, WA 6102, Australia \\
$^4$Research Center for Space and Cosmic Evolution, Ehime University, Matsuyama, Ehime 790-8577, Japan\\
$^5$Institute for Astronomy, University of Hawaii, 2680 Woodlawn Drive, Honolulu, HI 96822-1897, U.S.A. \\
}

\date{Accepted XXX. Received YYY; in original form ZZZ}

\pubyear{2024}

\begin{document}
\label{firstpage}
\pagerange{\pageref{firstpage}--\pageref{lastpage}}
\maketitle

% Abstract of the paper
\begin{abstract}
{\refbf Characteristic time scales in the stochastic UV-optical variability of quasars may depend on the mass of their black holes, $M_{\rm BH}$, as much as physical timescales in their accretion discs do. We calculate emission-weighted mean radii, $R_{\rm mean}$, and orbital timescales, $t_{\rm mean}$, of standard thin disc models for emission wavelengths $\lambda$ from 1\,000 to 10\,000~\AA , %\ to 1~$\mu$m, 
$M_{\rm BH}$ from $10^6$ to $10^{11}$~solar masses, and Eddington ratios from 0.01 to 1. At low $M_{\rm BH}$, we find the textbook behaviour of $t_{\rm mean}\propto M_{\rm BH}^{-1/2}$ alongside $R_{\rm mean} \approx \mathrm{const}$, but toward higher masses the growing event horizon imposes $R_{\rm mean} \propto M_{\rm BH}$ and thus a turnover into $t_{\rm mean}\propto M_{\rm BH}$. For quasars of $\log L_{\rm bol}=47$, the turnover mass, where $t_{\rm mean}$ starts rising is $M_{\rm BH}\approx 9.5$, which means that the turnover in $t_{\rm mean}$ is well within the range of high-luminosity quasar samples, whose variability time scales might thus show little mass dependence.
We fit smoothly broken power laws to the results and provide analytic convenience functions for $R_{\rm mean}(\lambda,M_{\rm BH},L_{3000})$ and $t_{\rm mean}(\lambda,M_{\rm BH},L_{3000})$. 
%in terms of the observables $\lambda$, $M_{\rm BH}$, and the monochromatic luminosity $L_{3000}$. 
%We then calculate variability structure functions for {\refbf $\sim$6\,200 of the} brightest quasars in the sky {\refbf at redshift $1<z<2.4$} with estimates for $M_{\rm BH}$ and $L_{3000}$, using lightcurves from NASA/ATLAS orange passband that span more than 7 years. The median luminosity of the accretion disc sample is $\log L_{\rm bol} / (\mathrm{erg}\, \mathrm{s}^{-1}) \approx {\refbf 46.5}$ 
}
%The data show at most a weak dependence of the variability {\refbf amplitude} on $M_{\rm BH}$ consistent with {\refbf either no mass dependence or with a turnover in} a model where the disc time scale drives variability amplitudes in the form $\log A/A_0 = 1/2\times \Delta t/t_{\rm orb}$. 
\end{abstract}


%%%%%%%%%%%%%%%%%%%%%%%%%%%%%%%%%%%%%%%%%%%%%%%%%%

%%%%%%%%%%%%%%%%% BODY OF PAPER %%%%%%%%%%%%%%%%%%

%\defcitealias{TWT}{Paper I}
%\defcitealias{incl}{Paper II}

\section{Introduction}\label{intro}

The emission from accretion discs in Active Galactic Nuclei (AGN) is variable on all time scales \citep[for reviews see][]{Ulrich97,Peterson01,Lawrence16}. Thus, it is routinely observed in all classes of AGN, where our view of the accretion disc is not obscured, and even used as a signature to identify AGN in time-domain sky surveys \citep[e.g.][]{Palanque11}. More importantly, characteristic behaviour within the seemingly stochastic variability is seen as a diagnostic tool to decipher physical properties of the discs {\refbf or of their central black holes}. %\citep{Lawrence16}. 
Sizes of accretion discs are probed with disc reverberation analysis \citep[e.g.][]{Sergeev05,Cackett07,Jiang17,Homayouni19,Yu20}, although there are also useful and complementary non-variability tools such as SED fitting \citep[e.g.][]{Malkan83,Laor_1990,Calderone_2013,Campitiello18,Lai23}. AGN accretion discs are also promising candidates for standardisable candles to extend studies of cosmology to the highest redshifts beyond the easy reach of other probes such as type-Ia supernovae. {\refbf These studies are based on relationships between disc luminosity and sizes of broad emission-line regions \citep[$R_{\rm BLR}-L$-relation,][]{Watson11,Khadka23}, or the X-ray to UV flux ratios \citep[$L_{\rm X}-L_{\rm UV}$-relation,][]{Risaliti19,Signorini23}, and thus} these studies would benefit from improved understanding of intrinsic disc properties.

Intriguingly, the physical origin of stochastic variability in AGN is not yet agreed upon; {\refbf suggestions include a variety of processes, from opacity-driven convection to a magnetic coupling between the hot X-ray corona and the cooler disc that dominates the energy output} \citep[e.g.][]{Jiang20,Sun20a,Neustadt22}, and thus it is not clear what behaviour to expect and how it relates to physical properties. A plausible candidate for intrinsic instabilities in the disc is turbulence from magneto-rotational instability \citep[MRI;][]{BH91}, although it is not yet established that this would predict the observed levels of variability in the integrated light of a whole disc. Separately, the disc is expected to respond to heating from a variable X-ray corona, although a limited energy budget suggests that this is not the principal origin of UV-optical variability in AGN discs \citep[e.g.][]{Uttley03,Arevalo08,Secunda24}. At present, we are far from a view of disc variability that is grounded in first-principles understanding and verifiable in numerical simulations, although attempts at the latter are getting ambitious \citep{Secunda24}, raising hope for future progress. 

On the observational side, current progress in the quest to identify mechanisms behind the variability centres on parametric descriptions of the stochastic behaviour, in the search for {\refbf dominant} parameters in a likely complex process \citep[e.g.][]{LawrPapa93,EdelNandra99,McHardy05}. Common descriptions of observed variability involve either the structure function (SF), most often for optical light curves \citep[e.g.][]{VB04,MacLeod10,Koz16}, or the power spectral density (PSD), most often for X-ray light curves \citep[e.g.][]{LawrPapa93,Paolillo23}, although opposite combinations exist as well \citep[e.g.][]{Arevalo24}. A common description of stochastic variability uses the damped random-walk paradigm \citep[e.g.][]{Kelly09,MacLeod10}, where specific interest is focused on the slope and amplitude of the SF or PSD, as well as breaks in slopes and their characteristic time scales. {\refbf While the DRW model posits a specific slope of the SF ($+1/2$) or PSD ($-2$) on timescales shorter than a decorrelation scale, observed deviations would hold clues about more complex behaviour, especially if they depended on physical parameters of the black hole and accretion disc.}

Initially, scaling behaviour of the X-ray PSD has been primarily related to black hole mass \citep[e.g.][]{LawrPapa93,EdelNandra99,McHardy05,Kelly13}; the optical behaviour in larger samples has been argued to be physically rooted in thermal fluctuations \citep{Kelly09}. On the UV-optical side, increasingly large and reliable data sets {\refbf have enabled many independent studies} \citep[see above, but also including][]{Zuo12,Mo14,Caplar17,Li18,Stone22,Arevalo24} {\refbf and developed} our view of scaling behaviour. For example, \citet{Burke21} suggest that a long-term damping time scale of the optical variability scales with black hole mass as well; \citet{TWT23} find that the rest-frame UV structure function is universal when clocks are run in units of thermal or orbital timescale that depends on wavelength and disc luminosity. \citet{Arevalo24} consider specifically the black hole mass dependence in the orbital timescale of UV emission. 

However, black hole mass estimates are still quite uncertain, and calculations of physical timescales in an accretion disc may be even less trusted as they are model-dependent. While a standard model for thin accretion discs exists \citep{SS73,NT73}, microlensing observations and disc reverberation experiments have suggested that the size scale of QSO discs may be {\refbf enlarged} by a factor of $\sim 3$ \citep[e.g.][]{Poo07,Mor10, Cack18}; however, the literature has not yet found agreement on mismatches of disc sizes with the standard model \citep[e.g.][]{Edel19,Yu20}, and {\refbf on their possible origin, which might relate to a larger and diffuse reprocessor \citep[e.g.][]{Fa16,McHardy18,Vinc21} or a more complex origin of signals} \citep[e.g.][]{Neustadt22,Secunda24}. {\refbf While size mismatches have initially questioned the viability of the thin-disc model, the additional reprocessors may help to reconcile the model with observations.}
%the measured signals} might still be correlated with disc size, such that the search for size ratios relative to the standard model may not question the model but rather constrain different mechanisms behind temperature waves travelling in the disc \citep{}.

When observed features are related to orbital or thermal timescales in the accretion discs, there are also slightly different approximating definitions used. Straightforward analytic equations are based on simple Newtonian forces in circular orbits and idealised gas properties \citep[for a handy summary in practical units, see e.g.][]{Kelly13}. Based on a universal temperature profile of $T(R)\propto R^{-3/4}$ in the outer parts of a standard disc and idealised black-body emission, analytic solutions were obtained that express the timescales as a function of bolometric luminosity $L_{\rm bol}$ and the restframe wavelength $\lambda_{\rm rest}$ of observed light; \citet{Mor10}, e.g., find an approximation for the disc scale length of $R \propto \lambda_{\rm rest}^{4/3} M_{\rm BH}^{2/3} (L_{\rm bol}/L_{\rm Edd})^{1/3}$, implying orbital and thermal time scales to follow $t \propto L_{\rm bol}^{1/2} \lambda_{\rm rest}^2$ independent of black hole mass. {\refbf Recently, \citet{Arevalo24} related their observations to the orbital timescale at the inner edge of the accretion disc, which is imposed by mass and spin of the black hole while being independent of the properties of the disc}. Clearly then, interpretations of scaling behaviour depend on approximations used in scale definitions, which is good reason for further investigation of what approximations work well in which part of parameter space.

Another question concerns which {\refbf observables are ideal when we look for scaling behaviour and parametrise accretion discs. Given a temperature gradient in accretion discs, we always expect properties to depend} on the observing wavelength. {\refbf But in terms of the fundamental parameters of the physical black hole and disc system, three different quantities are being used, of which only two are independent: black hole mass $M_{\rm BH}$, bolometric luminosity $L_{\rm bol}$ (which is expected to scale with mass accretion rate), and the Eddington ratio $R_{\rm Edd} = L_{\rm bol}/L_{\rm Edd}$ where $L_{\rm Edd}\propto M_{\rm BH}$.}
%the suggestion implicit in \citet{Mor10} is $L_{\rm bol}$, while many others prefer a combination of black hole mass $M_{\rm BH}$ and the Eddington ratio $R_{\rm Edd} = L_{\rm bol}/L_{\rm Edd} \propto L_{\rm bol}/M_{\rm BH}$. The latter combination has obviously an extra parameter to accommodate more complex behaviour, but {\refbf uses derived parameters rather than observables.} Whether $L_{\rm bol}$ or $R_{\rm Edd}$ should be given preference is less obvious -- while they are trivially related and thus seemingly interchangeable, the question is whether behaviour turns out to be independent of one but dependent on the other as a result of what is truly a causal dependence on black hole mass.
%
The use of these parameters in the analysis of real data is challenged by their large measurement uncertainties. Black hole mass is by far most often estimated from virial methods in single-epoch spectra, where it comes with an uncertainty of $\sim 0.5$~dex \citep{DallaBonta20,Bennert21}. {\refbf And particularly at highest luminosity, there is increasing evidence that virial mass estimates may be overestimated by 1~dex or more \citep{VLTI_J0920,VLTI_J0529}.}
$L_{\rm bol}$ is usually not observed but inferred from monochromatic luminosity with a standard bolometric correction (BC) that assumes that every AGN has the same spectrum \citep{Richards06b,Runnoe_2012}. While the UV-optical SEDs of most AGN appear largely uniform, it has been an obvious expectation that black holes of the largest mass will create the largest holes in the accretion discs and thus come with the coolest and reddest discs \citep{Laor_Davis_2011} that should have the smallest bolometric correction. Indeed, the most luminous QSOs appear to be powered by black holes with over $10^{10}$ solar masses and are consistent with BC factors that are $\sim 3\times$ lower {\refbf \citep[e.g.][]{Netzer19,Lai23,Wolf24}} than the standard values suggested for average QSOs \citep{Richards06b}. %\citet{Arevalo24} also explore trends in BC factors with black hole mass. 
Therefore, when standard BCs are used, $L_{\rm bol}$ will be biased by $M_{\rm BH}$. $R_{\rm Edd}$ is then a ratio obtained from a noisy $M_{\rm BH}$ and an $L_{\rm bol}$ estimate that is biased in the high-$M_{\rm BH}$ regime. 

{\refbf In this paper, we aim to use the most robust observables for parametrising accretion discs that we are aware of; we thus} work with observed luminosity directly instead of the noisier Eddington ratio and replace the mass-biased $L_{\rm bol}$ estimates with a more immediately observed monochromatic luminosity such as $L_{3000}$ or $L_{2500}$, where subscripts refer to wavelength in \AA ngstr\"om; either one is ideally inferred from spectral decomposition, with the former commonly published in QSO catalogues \citep[e.g.][]{Rak20} and the latter more often used in studies of X-ray-to-UV relations \citep[e.g.][]{Liu_2021}. This might seem like a small gain, given that an estimate of a monochromatic luminosity will depend not only on the accretion rate $\dot{M}$ of the black hole alone but also on the viewing angle of the non-isotropically emitting accretion disc, on any dust extinction by the AGN host galaxy or nuclear material, and also on the black hole spin. At least the spin dependence is lower than for $L_{\rm bol}$ \citep{Lai23} and the BC factor is removed, which depends on $M_{\rm BH}$ and $\dot{M}$. Further to that, the simple standard model ignores any Comptonisation of radiation from the inner disc and the complexities of photospheres in what will not be ideal thin discs. 
%While this may instill broad scepticism about taking any analysis of scaling behaviour too far, we may still optimistically attempt to quantify all possible effects in an ideal model and compare to observations. Here, it would be desirable if more capable and broadly accessible codes for calculating the spectra of thin and slim accretion discs such as \texttt{kerrbb} \citep{Li_2005} and \texttt{slimbh} \citep{Sadowski_2011, Straub_2011}, respectively, came with an option of outputting radially resolved emission profiles.

As we move into the era of big data on AGN variability, as facilitated by the Legacy Survey of Space and Time \citep[LSST;][]{Ivezic2008} starting soon at the Vera~C.~Rubin Observatory, we will wish to control for as many parameters in our interpretation of variability patterns, and ideally use a combination of variability and other diagnostics such as SED fitting \citep[e.g.][]{Laor_1990,Campitiello18,Lai23} and emission-line features \citep[e.g.][]{SH14,Marziani18,Mejia_Restrepo_2018} to enlarge the number of constraints on the physical parameters of black hole mass $M_{\rm BH}$, spin $a$, viewing angle $i$ and accretion rate $\dot{m}$, with a view to breaking remaining degeneracies, from which we currently suffer.

In this paper then, we investigate the {\refbf dependence of accretion} disc sizes and time scales {\refbf on black-hole mass and disc luminosity} and re-assess some of the choices made for their approximation. We will incorporate an approximate handling of General Relativity (GR) effects, and thus evaluate the dependence of time scales on the parameters $(L_{3000},M_{\rm BH},\lambda_{\rm rest},a)$. In Section~2, we describe our calculations of disc properties and choices of GR approximation. In Section~3, we present the results at face value and re-use analytic arguments to predict dominant simple approximations for the behaviour. As we confirm where the simple approximations apply, it will become clear that the influence of black hole mass depends heavily on the mass regime itself. The results are used in Section~4 to motivate a new parametrised approximation of the numerical grid, which can be used {\refbf conveniently} in future studies. In {\refbf a follow-up paper}, we will then investigate whether a mass effect can be empirically seen in the data of QSOs with high-mass black holes from {\refbf contemporary data} and to what extent it matches expectations worked out here.

\section{Accretion Discs}\label{sec:calc}

{\refbf Astrophysical accretion discs are a mature field, despite questions on how widely the most elegant solutions are applicable. The standard thin-disc model, proposed by \citet{SS73} and \citet{NT73}, has laid the foundation for describing discs around compact objects and is discussed in detail in modern textbooks devoted to the subject and developed over several editions \citep{FKR02,King23}. Although it has been shown to describe discs around stellar-mass bodies successfully, empirical confirmation of its applicability to AGN discs is lagging behind, partly because the evolutionary time scales in the latter are longer than the history of our exploration of AGN discs.

The simple model assumes that a geometrically thin disc orbits in a gravitational field that is completely dominated by the central black hole, which appears appropriate for discs around stellar-mass black holes but not necessarily for AGN \citep{Sirko03}. Neglecting any gravity from the disc mass itself substantially simplifies the path to a solution that describes the disc. The mass of the black hole uniquely determines the differential rotation profile of Keplerian orbits in the disc, where a steady state can be constrained by demanding the conservation of mass and angular momentum in a continuous accretion flow. This state implies a radial profile for the viscous heat release in the disc, which needs to be balanced by an equal loss of heat via thermal emission. With the further, plausible assumption that the disc is optically thick, this predicts a radial profile for the temperature of a disc surface that emits as a blackbody. Note that knowledge of the viscosity is not required to constrain the radial profile, although it is important for the vertical structure of the disc, which is ignored in this work as it does not affect the emitted spectrum to first order.
}

\subsection{Basic equations}

{\refbf We aim to calculate characteristic size scales and time scales of accretion discs for disc material that emits at different wavelengths. From the canonical thin-disc model introduced above,} we use the temperature profile together with black-body emission spectra to evaluate radial emission profiles for different wavelengths. Then we derive total disc luminosities as well as a light-weighted radius and light-weighted orbital and thermal time scales. 

We start with the standard temperature profile in Newtonian gravity
%of geometrically thin, optically thick high-viscosity discs, 
as specified by \citet{SS73}:
\begin{equation}\label{eq:T}
		T^4_{\rm N} (R) = \frac{3GM_{\rm BH}\dot{M}}{8\pi R^3\sigma}\left[1-\left(\frac{R_{\rm ISCO}}{R}\right)^{1/2}\right] ~,
\end{equation}
where $\sigma$ represents the Stefan-Boltzmann constant and $R$ denotes the radial distance from the centre. $R_{\rm ISCO}$ is the innermost stable circular orbit (ISCO) of the black hole as determined by the black hole spin. We calculate three cases for $r_{\rm ISCO}= R_{\rm ISCO}/R_{\rm S}$ with values of $(1.5,3,4.5)$, with the Schwarzschild radius $R_{\rm S}=2GM_{\rm BH}/c^2$; these correspond to spin values of $a=(+0.78,0,-1)$. 

We then apply approximate corrections for GR effects: for the emission spectrum, we follow the prescription of \citet{Hana89}, which combines gravitational redshift and time dilation effects into the modified temperature profile of
\begin{equation}\label{eq:GR_T}
	\begin{split}
		T_{\rm GR}(R) &=\sqrt{1-\frac{3}{2} \frac{R_{\rm S}}{R}}T_{\rm N}(R) \\
			&=\sqrt{1-\frac{3GM_{\rm BH}}{R c^2}}\left\{\frac{3GM_{\rm BH}\dot{M}}{8\pi R^3\sigma}\left[1-\left(\frac{R_{\rm ISCO}}{R}\right)^{1/2}\right]\right\}^{1/4} ~.
	\end{split}
\end{equation}

We choose to neglect the frame-dragging (Lense-Thirring) effect, since we are dealing with sizes much larger than the black hole ergosphere. We also %choose a face-on view of the disc when plotting radial profiles of the disc and 
ignore relativistic beaming effects, which become relevant near the inner edge of the disc. {\refbf Inclination dependence is ignored here and would require full GR ray tracing for an exact solution.} 

For any photon frequency $\nu$ and disc annulus at radius $R$ {\refbf and with a width of $dR$}, we create a radial annular flux density profile $F_\nu(R)$ as seen by an observer at luminosity distance $D$ by following \citet{SS73} in the form used by \citet{FKR02}:
\begin{equation}\label{eq:fnu}
    F_\nu (R) =\frac{4 \pi h\nu^3 \cos{i}}{c^2 D^2} \frac{R \mathrm{d}R}{e^{h\nu/k T(R)}-1} ~,
\end{equation}
where $h$ is the Planck constant and $k$ the Boltzmann constant. We then characterise the overall disc by calculating the bolometric luminosity as well as the monochromatic luminosity at $\lambda=3000$\AA , $L_{3000}$. We integrate over the range of inclination angles (with an average $\cos{i}$ factor of $1/2$) and the radial extend of the disc, using
\begin{equation}\label{eq:lwv}
	L_{3000} = 4 \pi D^2 \int_{R_{\rm ISCO}}^{R_{\rm out}} F_{3000}(R)  ~, % {\rm d}R,
\end{equation}
where $F_{3000}(R)$ represents $F_\nu (R)$ at $\lambda_{\rm rest}=3000$\AA \ and $R_{\rm out}$ denotes the outer edge of the disc. {\refbf Note, that the differential $dR$ is contained in the definition of $F_\nu(R)$.} Given the wavelength range of interest in this work beyond just a monochromatic luminosity, we generally choose $R_{\rm out}$ as the disc radius at $500$K, where we have surely captured the vast majority of thermal disc emission. In realistic AGN, we expect dust formation below temperatures of around 1\,000 to 1\,500~K, which means that the exact choice of outer cutoff for the disc will matter less than the complexity of real AGN and their deviation from ideal thin-disc models. From $L_{3000}$, we derive a fiducial {\it estimated} bolometric luminosity $L_{\rm bol,est} = f_{\rm BC}\times \lambda L_{3000}$ with $f_{\rm BC}=5.15$ %\times 0.75$ 
\citep{Richards06b} %,Runnoe_2012} 
as commonly done.

Separately, we calculate a {\it true} bolometric luminosity $L_{\rm bol}$, where we integrate the disc model $F_\nu (R)$ over relevant ranges in photon frequency to capture over 99\% of the thermal disc emission, using
\begin{equation}\label{eq:l_bol}
	L_{\rm bol} = 4\pi D^2 \int_{\nu_{\rm lo}}^{\nu_{\rm hi}}\int_{R_{\rm ISCO}}^{R_{\rm out}} F_\nu (R) {\rm d}\nu ~, \rm{with} ~ \cos i = 1/2 ~ \rm{again} ~,
\end{equation}
where $\nu_{\rm lo}$ and $\nu_{\rm hi}$ are frequencies corresponding to wavelength range of $\log(\lambda_{\rm rest}/$\AA $)= [2, 4.1]$. The Eddington ratio then follows from Equation~\ref{eq:l_bol} as: 
\begin{equation}\label{eq:r_edd}
	R_{\rm Edd} = L_{\rm bol}/L_{\rm Edd} 
\end{equation}
where the Eddington luminosity is 
\begin{equation}\label{eq:l_edd}
  L_{\rm Edd} = \frac{G M_{\rm BH} m_{\rm p} c}{\sigma_{\rm T}} ~,
\end{equation}
using the proton mass $m_{\rm p}$ and the Thomson scattering cross-section for the electron, $\sigma_{\rm T}$. We note that for $\log M_{\rm BH}=9$ and $\log R_{\rm Edd} = 0$ we find $\log L_{\rm bol}/(\mathrm{erg\,s^{-1}}) = 47.097$ and $\log L_{3000}/(\mathrm{erg\,s^{-1}}\,$\AA$^{-1}) = 42.908$; the difference of 4.189~dex is the factor $f_{\rm BC} \times 3\,000$~\AA .

From the radial emission profiles, we determine flux-weighted mean emission radii, $R_{\rm mean}$, for different wavelengths, assuming for simplicity a face-on view of the disc ($\cos{i}=1$) and thus using
\begin{equation}\label{eq:r_wei}
	R_{\rm mean} = \frac{\int_{R_{\rm ISCO}}^{R_{\rm out}} R F_\nu (R) }{\int_{R_{\rm ISCO}}^{R_{\rm out}} F_\nu (R) } ~.
	% {\rm d}R
    % = \frac{\int^{\text{\rtout}}_{\text{\risco}} \frac{R^2 {\rm d}R}{e^{\text{\hp} \nu/k T(R)}-1}}{\int^{\text{\rtout}}_{\text{\risco}} \frac{R {\rm d}R}{e^{\text{\hp} \nu/k T(R)}-1}}.
\end{equation}

Finally, we calculate a flux-weighted orbital time scale, $t_{\rm mean}$, for different wavelengths from the radial emission profile. Here, we start from a Newtonian definition of the orbital period, $t_{\rm orb,N}$, given by
\begin{equation}\label{eq:t_orbN}
  t_{\rm orb,N} = 2\pi \sqrt{\frac{R^3}{GM_{\rm BH}}} \simeq \gamma \times \left(\frac{M_{\rm BH}}{10^8 M_\odot}\right) 
     \left(\frac{R}{100 R_{\rm S}}\right)^{3/2}  ~ ~ \mathrm{days} 
\end{equation}
and add the GR time dilation effect with the modification
\begin{equation}
\label{eq:t_orb}
  t_{\rm orb} = t_{\rm orb,N} / \sqrt{1-\frac{3}{2} \frac{R_{\rm S}}{R}} ~.
\end{equation}
{\refbf Note that Eq.~4 in \citet{Kelly09} states a normalisation of $\gamma=104$~days, while we find $\gamma = \sqrt{32} \pi G/c^3 \times 10^{11}M_\odot =101.315$~days when using $G=6.6743\times 10^{-11}$~m$^3$kg$^{-1}$s$^{-2}$, $c=299\,792\,458$~m~s$^{-1}$ and $M_\odot=1.988475\times 10^{30}$~kg. While this is $\sim 2.6$\% shorter than the value from \citet{Kelly13}, it works out to be the same as their $t_{\rm orb}$ for $R\simeq 30 R_{\rm S}$ when GR effects are included}. The mean flux-weighted orbital time scale is thus
\begin{equation}
\label{eq:t_orb_m}
	t_{\rm mean} = \frac{\int_{R_{\rm ISCO}}^{R_{\rm out}} t_{\rm orb} F_\nu (R) }{\int_{R_{\rm ISCO}}^{R_{\rm out}} F_\nu (R) } ~. % {\rm d}R
\end{equation}

Since thermal time scales are just viscosity-dependent multiples of the orbital time scale \citep{FKR02}, we choose to proceed only with the more uniquely determined orbital time scale. 

Overall, we explore a disc parameter space that covers the parameter ranges of $\log (\lambda_{\rm rest}/$\AA $)=[3;4]$, $\log (M_{\rm BH}/M_\odot)=[6;11]$, and $\log (R_{\rm Edd})=[-2;0]$. %, all in steps of $0.1$~dex. 
In terms of black hole spins, we explore three values, $a=-1$ (maximum retrograde spin), $a=0$ (Schwarzschild black hole), and $a=+0.78$ (a high prograde spin).

These calculations determine the isotropically averaged luminosity of an accretion disc, while the luminosity measured for observed discs will depend on inclination. We also ignore inclination-dependent GR effects, which lead to second-order modifications of the spectral shape, the observed luminosity, and the mean scales.
%{\bf (CHECK whether L3000 is face-on isotropic estimate, if yes, simply apply isotropic correction factor, the mean cos i, i.e. 2/pi, to the tables and plots? Also, fix 3,4,5, final check 8 and 11)}

\begin{figure}%[!htb]
\begin{center}
\includegraphics[width=\columnwidth]{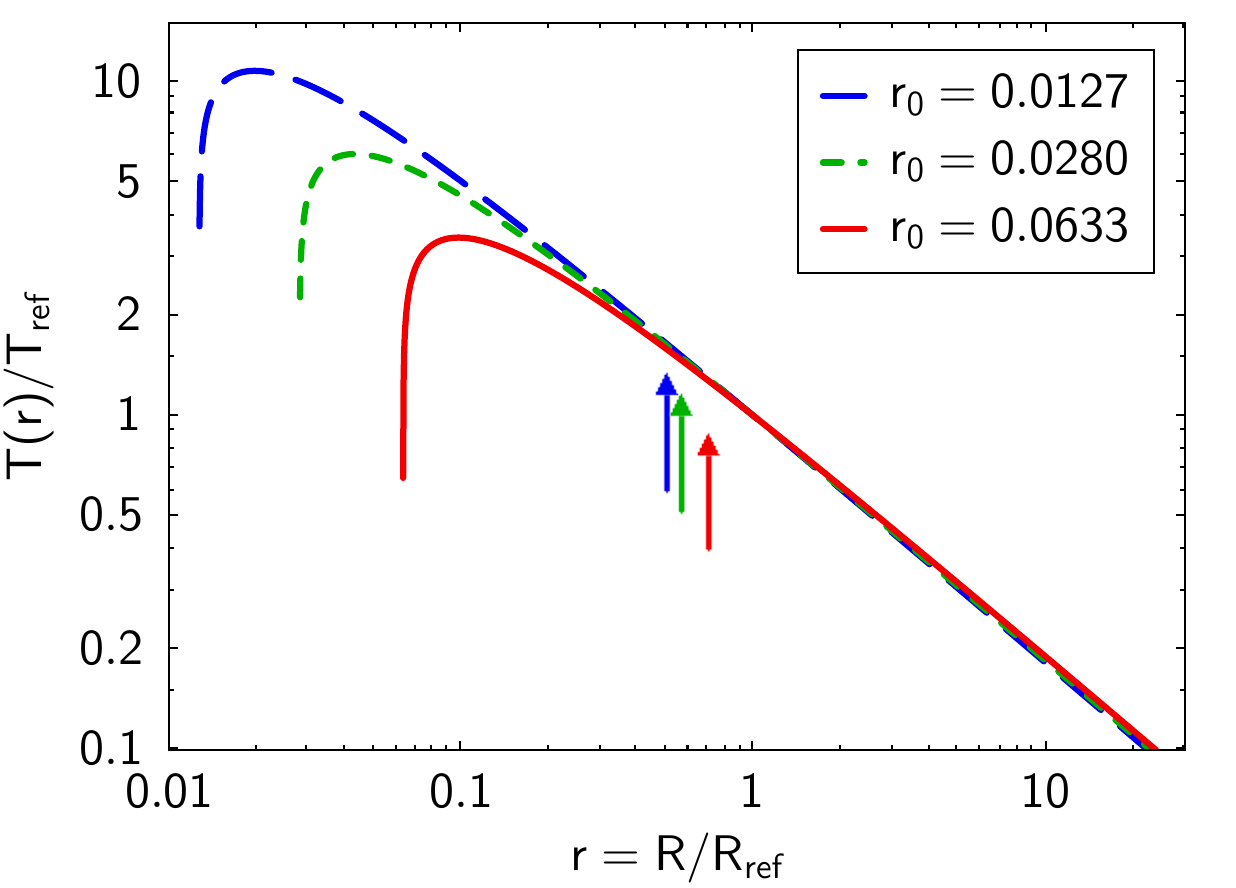}
\caption[]{\refbf Disc temperature profiles derived from Eq.~\ref{2Dfamily} using $T_{\rm ref}=10^4$~K. For a spin-free black hole of $\log (M_{\rm BH}/M_\odot)=9$, which has $\log (R_{\rm ISCO}/\mathrm{m})=12.947$, $\log (L_{\rm Edd}/(\mathrm{erg~s^{-1}}))=47.10$ and $\dot{M}_{\rm Edd}=22.2M_\odot~\mathrm{yr}^{-1}$, the three curves represent the cases $R_{\rm Edd}=(1,0.1,0.01)$ in order of increasing $r_0$. The given $r_0$ values correspond to $R_{\rm 10^4\mathrm{K}}/R_{\rm S}\simeq (236,107,47)$. While the curves are universal for chosen values of $T_{\rm ref}$ and $r_0$, $R_{\rm Edd} \propto M_{\rm BH}$ at any fixed $r_0$, and thus e.g., for $R_{\rm Edd}=0.1$ the curves represent $\log (M_{\rm BH}/M_\odot)=(8,9,10)$. The arrows mark the mean emission radii $R_{\rm mean}$ for $\log (\lambda/$\AA $)=3.5$, which are always $>R_{\rm ISCO}$ but converge to a constant for vanishing $R_{\rm ISCO}$ (or $r_0$).
%$R_{\rm Edd}=(10,1,0.1)$. 
%At that black hole mass a disc with $R_{\rm Edd}=0.1$ reaches a maximum temperature of just over 30\,000~K.
}\label{fig:Tprofile}
\end{center}
\end{figure}
% Found from JJ R1M cubes:  log R_10K (1e9 MBH, REdd=1, 0.1, 0.01) = 1.1648, 0.8186, 0.4616, which can be compared to mean log lamAA=3.5 emission point at log RWE () = 0.8723, 0.5740, 0.3085, so now add arrows to figure! So, for the figure the log RWE/R10k () = -0.2925,-0.2446,-0.1531

\subsection{Building an intuition}\label{sec:intuition}

To first order, the discs in our model are black-body emitters with a common temperature profile {\refbf where larger discs are more luminous, simply due to a larger surface area for any temperature interval; two further factors impact the profile, mostly in its innermost part, which are the size of the hole in the disc due to the ISCO and the gravitational redshift. If we define a size scale $R_{\rm ref}$ for a fixed temperature $T_{\rm ref}$ as a primary ordering parameter, then} the complete family of temperature profiles in Eq.~\ref{eq:GR_T} is given by
\begin{equation}\label{fullTR}
    \frac{T^4}{T^4_{\rm ref}} = \frac{R^3_{\rm ref}}{R^3} 
            \frac{1-\sqrt{\frac{R_{\rm ISCO}}{R}}}{1-\sqrt{\frac{R_{\rm ISCO}}{R_{\rm ref}}}}  
            \frac{\left(1-\frac{3}{2}\frac{R_{\rm S}}{R}\right)^2}{\left(1-\frac{3}{2}\frac{R_{\rm S}}{R_{\rm ref}}\right)^2}    ~.
\end{equation}

{\refbf Intuitive predictions for how the total disc luminosity $L$ scales with the disc scale $R_{\rm ref}$ may use the approximation $T\propto (R/R_{\rm ref})^{-3/4}$ and still find two different answers, depending on which integration limits $R_{\rm in}$, $R_{\rm out}$ are used: assuming identical limits, a change in $R_{\rm ref}$ will change the emitted luminosity in each annulus formed by a fixed radius interval $[R;R+dR]$ by $L\propto T^4 \propto R_{\rm ref}^3$; in contrast, scaling the integration limits such that $R_{\rm in}/R_{\rm ref}=\rm const$ and $R_{\rm out}/R_{\rm ref}=\rm const$ will change emitted luminosity in each annulus formed by a fixed temperature interval $[T;T+dT]$ by as much as its area and thus by $L \propto R_{\rm ref} dR/dT\propto R_{\rm ref}^2$, which emphasises the role of the ISCO. 

Working with Eq.~\ref{fullTR} we explore the exact properties of this family of profiles by changing variables to $r=R/R_{\rm ref}$ (such that $T(r=1)=T_{\rm ref}$), $r_0=R_{\rm ISCO}/R_{\rm ref}$, and we also use $r_{\rm ISCO}=R_{\rm ISCO}/R_{\rm S}$ as defined above. We thus get
\begin{equation}
    \frac{T^4}{T^4_{\rm ref}} = r^{-3} 
            \frac{1-\sqrt{r_0/r}}{1-\sqrt{r_0}}  
            \frac{\left(r_{\rm ISCO}-\frac{3}{2}r_0/r\right)^2}{\left(r_{\rm ISCO}-\frac{3}{2}r_0\right)^2 }  ~.
\end{equation}
Given that we only have $r_{\rm ISCO}\ge 3/2$ and $r_0<1$, this equation is well-defined for any place in the disc (i.e., at $r\ge r_0$). The family of profiles is spanned by three parameters: a primary size scale $R_{\rm ref}$, the relative size of the hole in the disc caused by the ISCO (which depends on mass and spin of the black hole), and a relativistic correction (which depends only on mass of the black hole). For a fixed black hole spin, this simplifies to a 2-parameter family; e.g., in the case of a non-rotating black hole ($r_{\rm ISCO}=3$), we get
\begin{equation}\label{2Dfamily}
    \frac{T^4}{T^4_{\rm ref}} = r^{-3} 
            \frac{1-\sqrt{r_0/r}}{1-\sqrt{r_0}}  
            \left(\frac{2-r_0/r}{2-r_0}\right)^2  ~,
\end{equation}
which is a 2-parameter family spanned by $R_{\rm ref}$ and $r_0$. Figure~\ref{fig:Tprofile} shows temperature profiles over the normalised radial coordinate $r$ for three example values of $r_0$. How $r_0$ relates to black hole mass and Eddington ratio will be worked out in the following steps.

Obviously, for a fixed value of $r_0$, only a 1-parameter family remains, which consists of identical temperature profiles that are simply scaled by $R_{\rm ref}$. A fixed $r_0$ means that} the inner edge of the disc at $R_{\rm ISCO}$ and the outer edge $R_{\rm out}$ scale linearly with $R_{\rm ref}$. If we define the outer edge by a fixed temperature $T_{\rm out}$, where the disc ceases to contribute to the UV-optical emission due to low temperature, then $R_{\rm out}/R_{\rm ref}$ will be automatically constant given the fixed $T(R/R_{\rm ref})$ profile. Requiring $R_{\rm ISCO}/R_{\rm ref}=\mathrm{const}$ demands $M_{\rm BH} \propto R_{\rm ref}$. This 1-parameter family has {\refbf temperature profiles of identical shape, apart from an overall radial scale. 

This family also emits spectra of identical shape, apart from an overall luminosity scale. This can be seen by inserting a fixed scaled temperature profile into the luminosity integral given in Equations~\ref{eq:fnu} and \ref{eq:lwv}: as long as the emitted spectrum per unit surface area depends only on the surface temperature in a function $g_\nu(T)$, as is the case for blackbody radiation, we can write the luminosity integral as
\begin{equation}\label{Linteg}
    L_\nu = 4\pi \int g_\nu(T(R)) R dR  ~.
\end{equation}
When using a fixed $T(r) = T(R/R_{\rm ref})$, we get a fixed $g_\nu(r)$, and a change of variable using $R=r\times R_{\rm ref}$ and $dR = dr \times R_{\rm ref}$ leads to
\begin{equation}\label{Linteg2}
    L_\nu = 4\pi \int g_\nu(r) R dR = 4\pi R^2_{\rm ref} \int g_\nu(r) r dr ~.
\end{equation}
Hence, the mean surface luminosity at any wavelength is constant in a 1-parameter family with fixed $r_0$, and the monochromatic and bolometric luminosities scale as $L\propto R_{\rm ref}^2$. This family of discs has evidently constant values of $R_{\rm ref}/\sqrt{L_{3000}}$ and $R_{\rm ref}/\sqrt{L_{\rm bol}}$. A fixed $r_0=\rm const$ implies $M_{\rm BH}\propto R_{\rm ref}$ and hence this family also has constant values of $M_{\rm BH}/\sqrt{L_{3000}}$. Furthermore, the fixed shape of the emission profile implies constant values of $R_{\rm mean}/R_{\rm ref}$ and thus constant values of $R_{\rm mean}/\sqrt{L_{3000}}$. While the curves in Figure~\ref{fig:Tprofile} are universal for any chosen $T_{\rm ref}$ and $r_0$ values, the relation $r_0 \propto M_{\rm BH}/\sqrt{L_{\rm bol}}$ at fixed $r_0$ implies also $R_{\rm Edd} \propto L_{\rm bol}/M_{\rm BH} \propto M_{\rm BH}$. Higher-mass discs therefore reach lower maximum temperatures, even for significant Eddington ratios \citep[see also][]{Laor_Davis_2011}.

%For $\log (M_{\rm BH}/M_\odot)=10$, the curves represent $R_{\rm Edd}=(10,1,0.1)$. At that black hole mass an accretion disc with $R_{\rm Edd}=0.1$ reaches a maximum temperature of just over 30\,000~K.

For the orbital timescales in this family of profiles we find, after changing from the variable $R$ to $r=R/R_{\rm ref}$ in Equations~\ref{eq:t_orb} and \ref{eq:t_orb_m},
\begin{equation}\label{torb_scaled}
    t_{\rm orb}(r)
    = 2\pi \left. \sqrt{\frac{r^3 R^3_{\rm ref}}{GM_{\rm BH}}} \middle/ \sqrt{1-\frac{3}{2}\frac{r_0}{r}\frac{1}{r_{\rm ISCO}}} \right. ~,
\end{equation}
which simplifies in a family of fixed $r_0$ and fixed $r_{\rm ISCO}$ and with $M_{\rm BH}\propto R_{\rm ref}$ to
\begin{equation}\label{torb_scaled_f}
    t_{\rm orb}(r)/R_{\rm ref} = f(r) ~,
\end{equation}
so that all these discs have constant values of $t_{\rm mean}/\sqrt{L_{3000}}$ as well.}

The second parameter, $r_0=R_{\rm ISCO}/R_{\rm ref}$, covers the variation of the inner disc edge, which %acts as an integration limit and also 
affects the inner temperature and emission profile and thus the spectral energy distribution, the mean surface luminosity and bolometric correction. It also varies the inner edge of the disc integration and flux-weighted averaging, and instead of the intuitive $R_{\rm ISCO}/R_{\rm ref}$, the second parameter could be chosen to be interchangeably $R_{\rm ref}/\sqrt{L_{3000}}$, $M_{\rm BH}/\sqrt{L_{3000}}$, or $R_{\rm mean}/\sqrt{L_{3000}}$, {\refbf because all of them vary strictly monotonically with $r_0$}. Note that $t_{\rm mean}/\sqrt{L_{3000}}$ {\refbf does not vary monotonically with $r_0$ and thus could not be a unique second parameter; this is because} $t_{\rm mean}$ not only depends on the integration limits set by $R_{\rm ISCO}$ and thus on a combination of $M_{\rm BH}$ and black hole spin $a$, but additionally depends on $M_{\rm BH}$ itself via the Keplerian orbits. 
{\refbf In Figure~\ref{fig:Tprofile}, flux-weighted mean emission radii $R_{\rm mean}$ are marked with arrows (for $\log (\lambda_{\rm rest}/$\AA $)=3.5$). The three cases of $r_0$ hint at the fact that a variation of $r_0$ hardly affects $R_{\rm mean}$ as long as $r_0$ is small; but for large values of $r_0$ the inner disc edge at $R_{\rm ISCO}$ can be larger than $R_{\rm mean}$ for small $r_0$. As $r_0$ grows, it reaches a regime, where it affects $R_{\rm mean}$ strongly, pushing it outwards, as we shall see in detail in Section~\ref{sec:results}. }

\begin{figure*}%[!htb]
\begin{center}
\hspace{0mm}
\includegraphics[width=0.81\columnwidth]{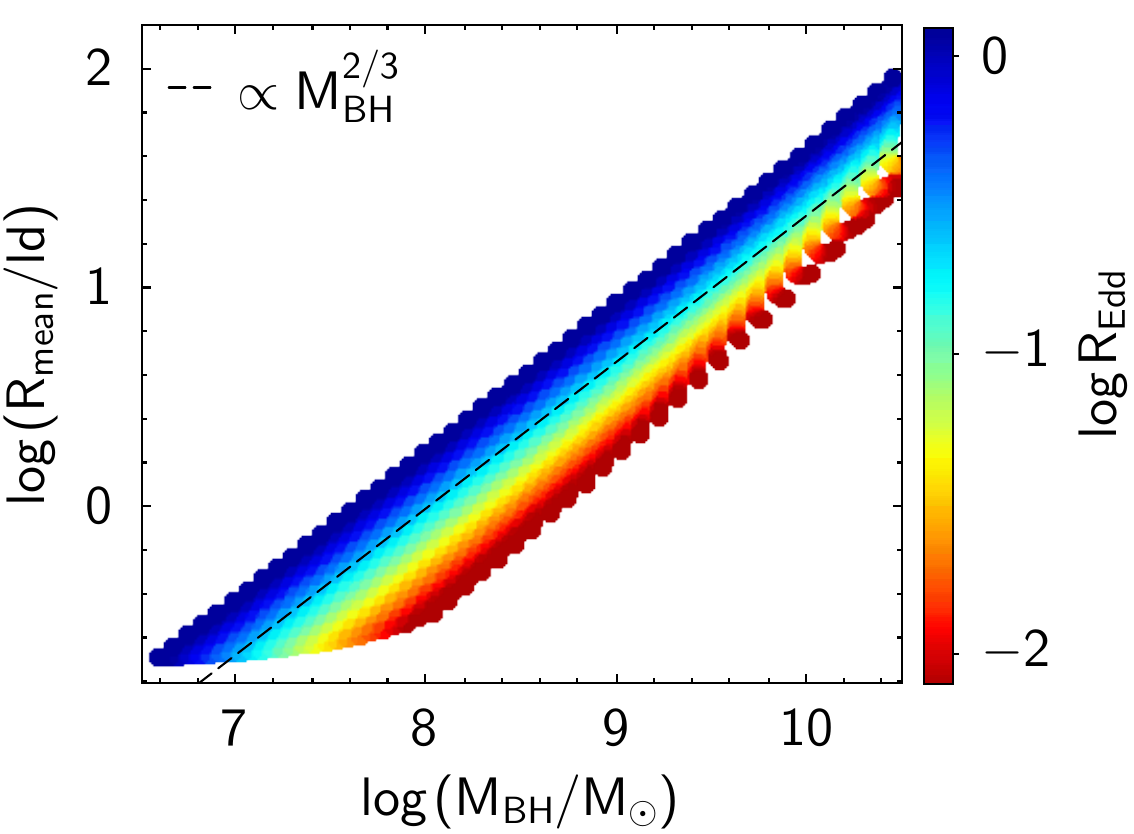}
\hspace{4mm}
\includegraphics[width=0.81\columnwidth]{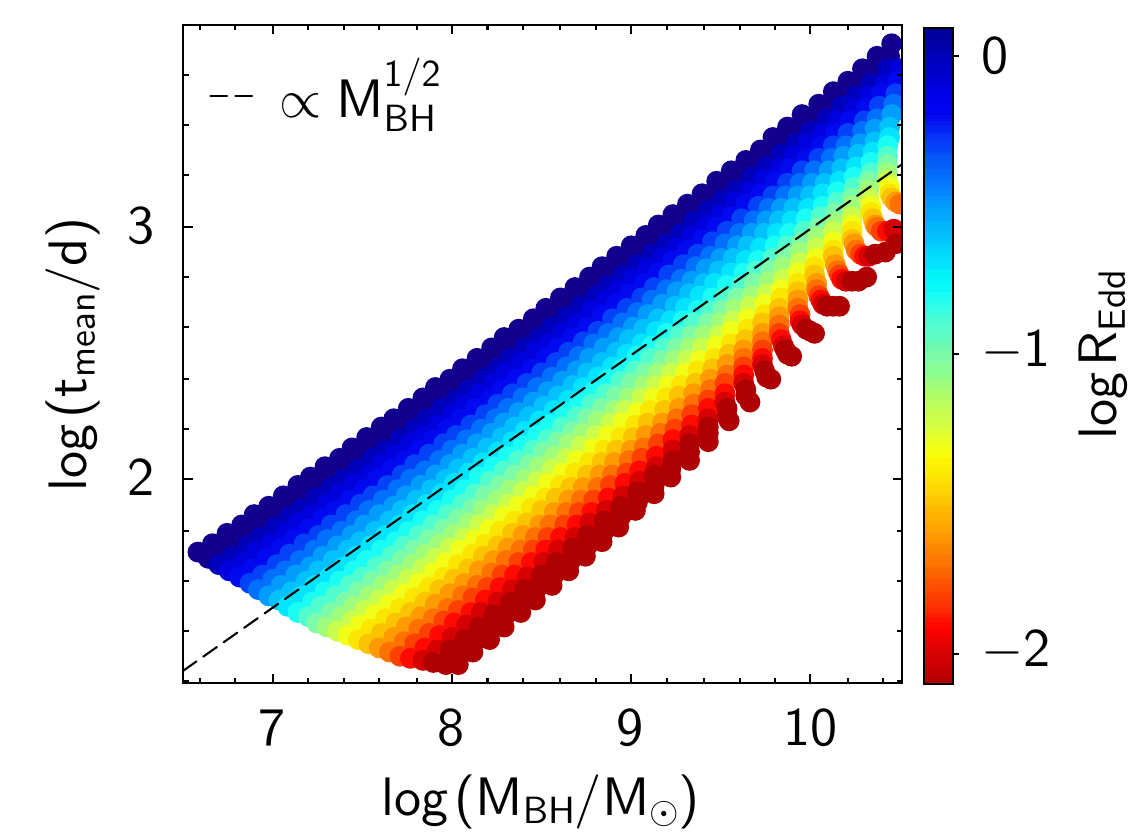} \\
\includegraphics[width=0.79\columnwidth]{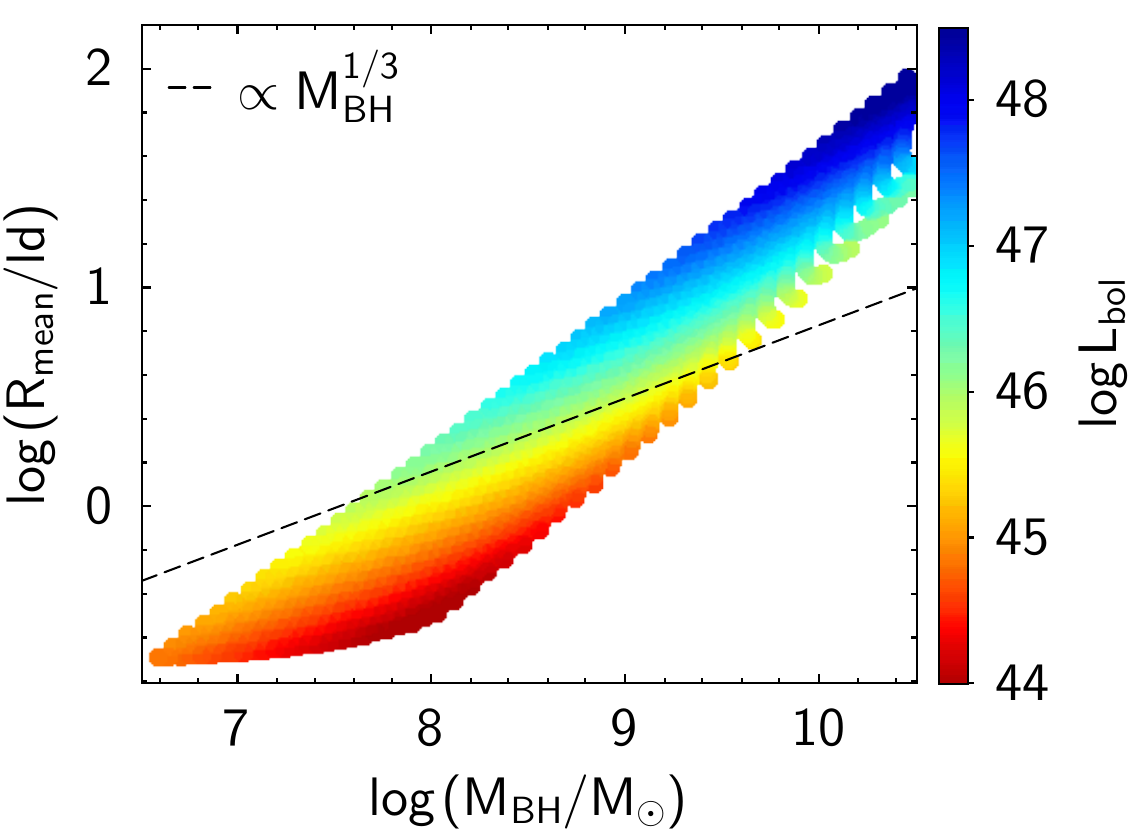}
\hspace{5mm}
\includegraphics[width=0.79\columnwidth]{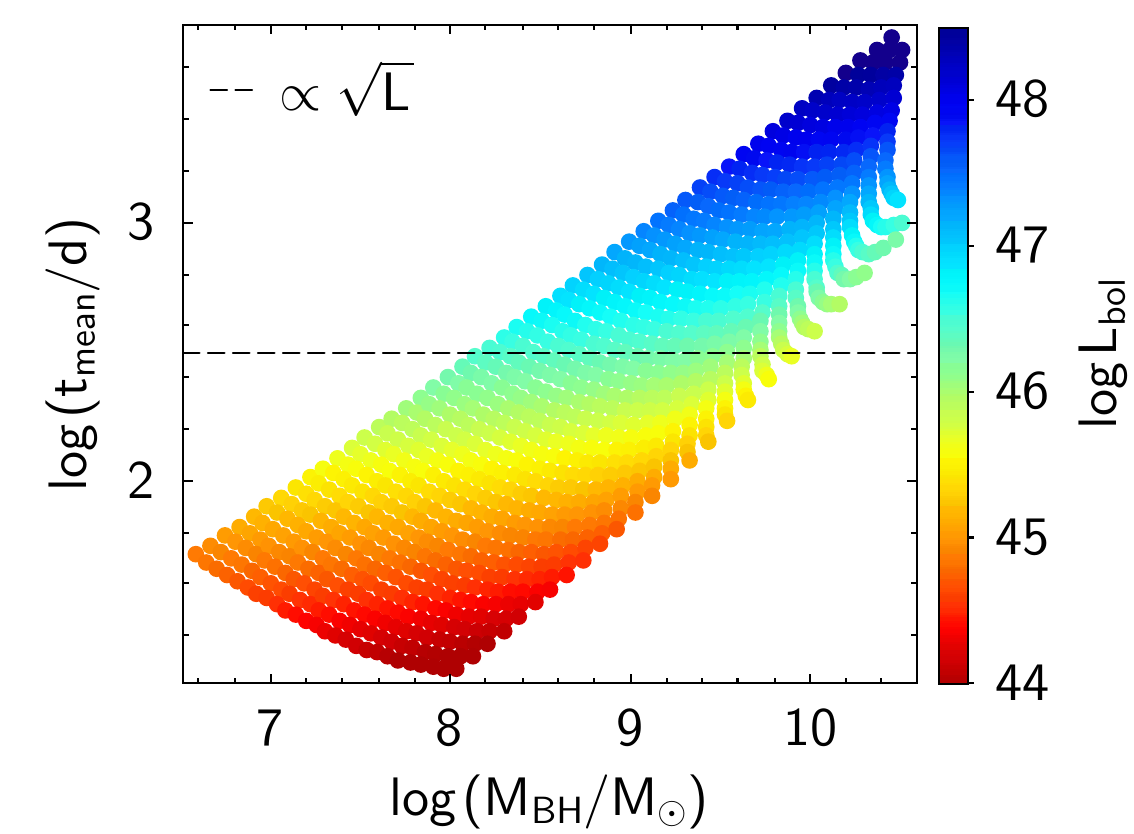} \\
\includegraphics[width=0.79\columnwidth]{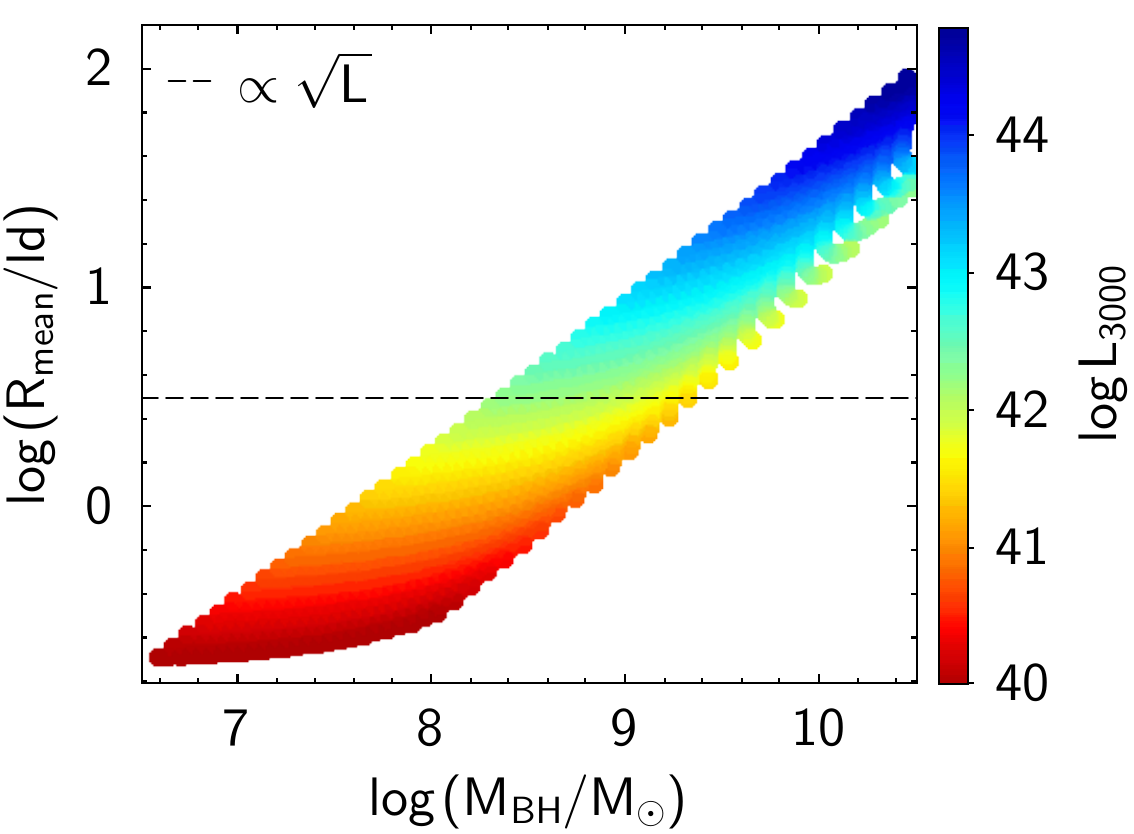} 
\hspace{5mm}
\includegraphics[width=0.79\columnwidth]{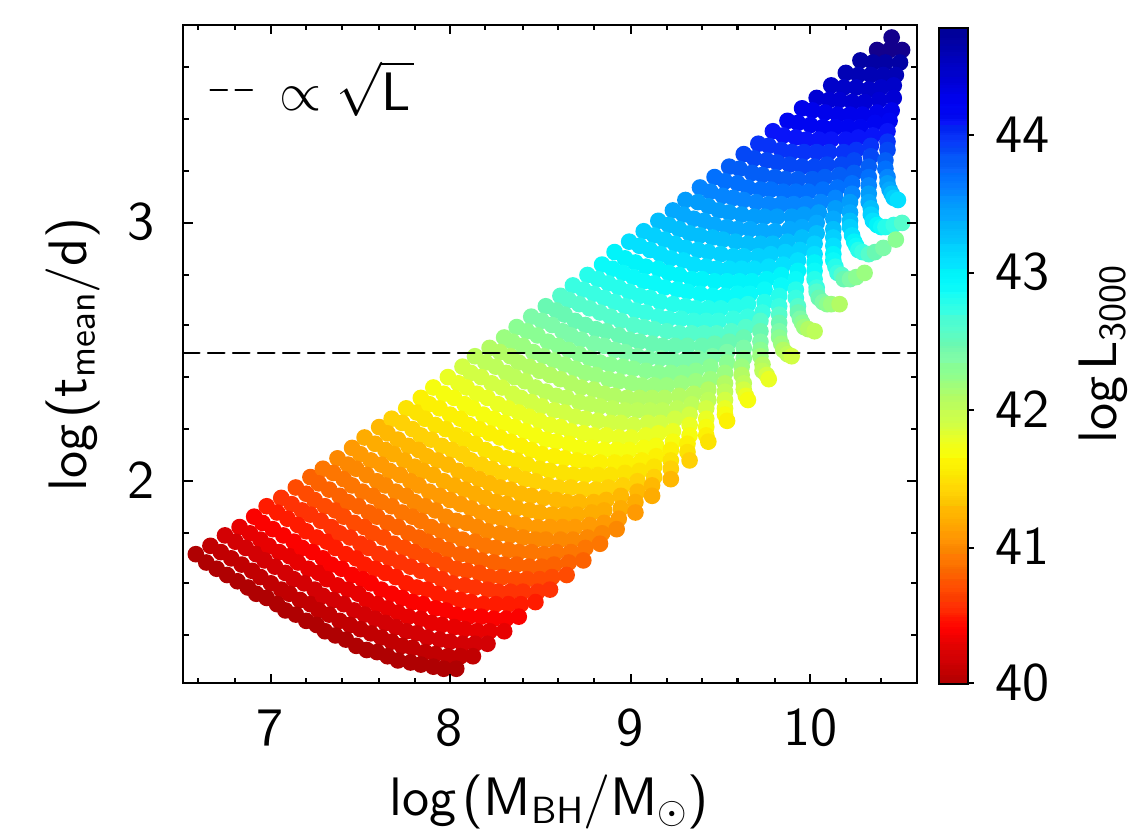} \\
\caption[]{Flux-weighted size scale (mean emission radius) $R_{\rm mean}$ ({\refbf left}) and orbital time scale $t_{\rm mean}$ ({\refbf right}) at {\refbf$\log (\lambda /\textup{\AA})=3.5$} for a range of accretion discs (GR approximation with spin $a=0$) using three colour codes: Eddington ratio $R_{\rm Edd}$ ({\refbf top}), true bolometric luminosity $L_{\rm bol}$ (centre), and monochromatic luminosity $L_{3000}$ ({\refbf bottom}), which often acts as a proxy for $L_{\rm bol}$. Dashed lines show the {\refbf power law scaling suggested by the approximation in Eq.~\ref{eq:r_ana_dep}}.
%, and the solid line is from \citet{Arevalo24}. 
}\label{fig:rM_tM_Redd}
\end{center}
\end{figure*}

\subsection{Previous parametrisations for $R_{\rm ref}$ and $t_{\rm orb}$}
{\refbf
\citet{Mor10} observed a small number of microlensed quasars with the intent of measuring the sizes of their accretion discs and comparing it to the standard thin-disc model. Following \citet{FKR02}, they derived a size scaling with emitted wavelength using the Wien displacement law using the radius $R_{\rm M10}$ at which the disc temperature matches the wavelength $kT_{\lambda_{\rm rest}}=hc/\lambda_{\rm rest}$, where $k$ is the Boltzmann constant and $h$ is the Planck constant, and found
\begin{equation}\label{eq:r_ana}
\begin{split}
    R_{\rm M10} &=\left( \frac{45G\lambda_{\rm rest}^4 M_{\rm BH}\dot{M}}{16\pi^6hc^2}\right)^{1/3} \\
    &=9.7\times10^{15}\left(\frac{\lambda_{\rm rest}}{\mu {\rm m}}\right)^{4/3}\left(\frac{M_{\rm BH}}{10^9 M_\odot}\right)^{2/3}\left(\frac{L_{\rm bol}}{\eta L_{\rm Edd}}\right)^{1/3} {\rm cm} ~,
\end{split}
\end{equation}
where $G$ is the gravitational constant and $\dot{M}$ the accretion rate. The fraction of the rest mass energy of the accreted matter that is converted into emitted radiation, $\eta=L_{\rm bol}/(\dot{M}c^2)$, is known as the radiative efficiency and is estimated to be 0.057 for spin $a=0$, but larger for co-rotating black holes and smaller for counter-rotating ones. Given $L_{\rm Edd}\propto M_{\rm BH}$, this is a scaling of
\begin{equation}\label{eq:r_ana_dep}
     R_{\rm M10}\propto M_{\rm BH}^{2/3} R_{\rm Edd}^{1/3} \propto M_{\rm BH}^{1/3} L_{\rm bol}^{1/3} ~.
\end{equation} 
At fixed $r_0$, this agrees with our findings: a fixed $r_0$ implies a constant temperature profile except for scaling by $R_{\rm ref}$ and thus predicts $R_{\rm ref}\propto \sqrt{L_{3000}} \propto \sqrt{L_{\rm bol}}$. Starting from equation~\ref{eq:r_ana_dep} and inserting the relation $M_{\rm BH} \propto \sqrt{L_{3000}} \propto \sqrt{L_{\rm bol}}$ (imposed by the fixed $r_0$) similarly yields
\begin{equation}\label{eq:r_ana_dep2}
     R_{{\rm M10,fixed-}r_0}\propto L_{\rm bol}^{1/6} L_{\rm bol}^{1/3} \propto \sqrt{L_{\rm bol}} ~.
\end{equation} 
However, when varying $r_0$ the derivation by M10 will be different from ours, because changing $r_0$ changes the shape of the temperature profile so that $L_{3000}$ and $L_{\rm bol}$ are no longer proportional.

Observations of a disc size scaling with black hole mass depend on trends of Eddington ratio with black hole mass in the observed sample. \citet{Mor10} found an empirical scaling of $R_{\rm mean}\propto M_{\rm BH}^{0.8\pm0.17}$ from a small sample, which works with the scaling of thin discs if $R_{\rm Edd} \propto M_{\rm BH}^{0.4\pm0.5}$ on average; this is not in conflict with studies of the bulk quasar population \citep[e.g.,][]{Shen08}.

We can then find the orbital timescale in Eq.~\ref{eq:t_orbN} at the scale radius $R_{\rm M10}$ \citep[as shown, e.g., by][]{MacLeod10} and find a scaling of $t_{\rm orb,M10} \propto \sqrt{M_{\rm BH} R_{\rm Edd}} \propto \sqrt{L_{\rm bol}}$. The orbital timescale thus depends only on the bolometric luminosity, without an additional dependence on the black hole mass. This approximation has often been used in studies of the variability structure function, from \citet{MacLeod10} to \citet{TWT23}. Typically, $L_{\rm bol}$ is not observed, but derived from a monochromatic luminosity $L_\lambda$ with a fixed bolometric correction that is typical for a mean quasar SED; in the thin-disc model, however, the bolometric correction depends on $r_0$, i.e., on the ratio $M_{\rm BH}/L_\lambda$. The following section presents our numerical results.}

\section{Model results}\label{sec:results}

We first use the single wavelength of {\refbf $\log (\lambda /\textup{\AA})=3.5$} from the grid of discs without black hole spin to explore the dependence of the light-weighted radius scale $R_{\rm mean}$ and orbital time scale $t_{\rm mean}$ on black hole mass, luminosity, and Eddington ratio; we will also differentiate between the true $\lambda$-integrated bolometric luminosity $L_{\rm bol}$ and the monochromatic luminosity $L_{3000}$ that is a common proxy for the bolometric luminosity through simple scaling with a BC factor. Specifically by looking at the mass dependence of the disc size and orbital time scales at fixed luminosity, we will find that it follows not one power law but a smoothly broken power law as the driving factor for the scale changes from low mass to high mass. We will compare the results from the numerical grid with simple analytic approximations and then {\refbf proceed to develop an improved approximation in Section~\ref{sec:approx}}.

%\begin{figure}%[!htb]
%\begin{center}
%\includegraphics[width=\columnwidth]{M_RRM10_lambda.pdf}
%\caption[]{Ratio of flux-weighted size scale (mean emission radius) $R_{\rm mean}$ to the size approximation in \citet{Mor10}, $R_{\rm M10}$, colour-coded by emission wavelength $\lambda$ (GR approximation with spin $a=0$). For optical wavelengths, the flux-weighted sizes are $\sim 5\times$ larger than $R_{\rm M10}$, but the scales of UV emission may be enlarged by the large $R_{\rm ISCO}$ around the most massive black holes. Four groups of discs are shown, with $\log L_{3000}/(\mathrm{erg\,s^{-1}}\,$\AA$^{-1})$ values of 40, 41, 42 and 43; the black hole mass range is driven by the Eddington ratio limits of the calculated grid. }\label{fig:t_M_lamM10}
%\end{center}
%\end{figure}

\subsection{Simple scaling approximations for size and time scales}

In Figure~\ref{fig:rM_tM_Redd}, we show the mass dependence of the size scale ({\refbf left column}) and the orbital time scale ({\refbf right column}) while colour-coding the discs with Eddington ratios ({\refbf top row}) and luminosities (centre and {\refbf bottom rows}). We see the 2-parameter family of discs squeezed into a narrow distribution of size scales proportional to black hole mass, such that a power law index can easily by fitted. The approximation {\refbf in Eq.~\ref{eq:r_ana_dep} } predicts $R_{\rm mean}\propto M_{\rm BH}^{2/3}$ at fixed Eddington ratio and $R_{\rm mean}\propto M_{\rm BH}^{1/3}$ at fixed luminosity $L_{\rm bol}$. A slope of $2/3$ (dashed line, top left panel) fits the general trend. 
%How an observed sample of QSOs will behave on average, depends on trends of Eddington ratio with black hole mass. From observations of a small number of microlensed quasars, \citet{Mor10} found $R_{\rm mean}\propto M^{0.8}$ (although, e.g., \citet{Shen08} present a more elaborate study).
{\refbf The orbital time scale behaviour of} $t_{\rm orb} \propto \sqrt{M_{\rm BH}}$ at fixed Eddington ratio {\refbf roughly fits the trend as well (dashed line, top right} panel). 
%as does the fit to similar calculations by \citet{Arevalo24} who found $t_{\rm orb}\propto M^{0.65}$ (solid line, bottom left panel). 
The spread {\refbf in time scales} is wider than that {\refbf in} size scales (at fixed mass $t_{\rm orb} \propto R^{3/2}$) and a curvature beyond a single power law is more noticeable. While a second parameter could be captured in a scaling with Eddington ratio, the slope of a power-law fit to the latter depends on black hole mass. %(see Figure~\ref{fig:t_R_M}). 
%A mean slope at fixed mass of $t_{\rm orb} \propto R_{\rm Edd}^{1/3}$ as predicted by the \citet{Mor10} approximation is similar to the \citet{Arevalo24} fit of $t_{\rm orb}\propto R_{\rm Edd}^{0.35}$.
%however, either slope applies best at masses of $9<\log M_{\rm BH}<10$, while the slope exceeds $1/2$ at masses below $\log M_{\rm BH}<7.x$ and conversely becomes nearly mass-independent at $\log M_{\rm BH}>10$.

The {\refbf centre row} of Figure~\ref{fig:rM_tM_Redd} renders the same points colour-coded by luminosity $L_{\rm bol}$. The mass dependence of the scales is generally weaker when evaluated at fixed $L$ rather than fixed $R_{\rm Edd} \propto L/M_{\rm BH}$ due to the intrinsic additional factor $M_{\rm BH}$. Dashed lines {\refbf show the slope} $1/3$ ({\refbf centre left} panel) {\refbf from Eq.~\ref{eq:r_ana_dep} and the mass independence of the orbital timescale at fixed $L_{\rm bol}$} ({\refbf centre right} panel); {\refbf these} are now meant to follow the distribution of points in one colour, not the overall distribution.
%The \citet{Arevalo24} fit translates int $t_{\rm orb}\propto M_{\rm BH}^{0.3}$, but the \citet{Mor10} solution seems at least as acceptable. 
However, the evident relation is still not a single power law, but shows curved behaviour. We note that this is the true $L_{\rm bol}$ as determined by integrating over all emission; in practice, $L_{\rm bol}$ is often estimated from a monochromatic luminosity with a standard mass-independent bolometric correction factor. Thus, we consider a relation between $t_{\rm mean}$ and $L_{3000}$ next.

The {\refbf bottom row} of Figure~\ref{fig:rM_tM_Redd} renders the same points colour-coded by luminosity $L_{3000}$, which we prefer as a more robust observable when considering a wide range of black hole masses. %While the mass dependence is generally even weaker for $L_{3000}$ than for $L_{\rm bol}$, 
{\refbf At fixed $L_{3000}$ the numerically calculated $R_{\rm mean}$ depends only weakly on the black hole mass. The time scale} $t_{\rm mean}$ declines with increasing mass at fixed $L_{3000}$ in the low-mass regime, then shows a parabolic turnover at intermediate mass, and finally increases with mass at high black hole masses. The mass-independent scaling {\refbf of $t_{\rm orb} \propto \sqrt{L_{\rm bol}}$} will capture the average scales throughout the turnaround at intermediate masses of $\log M_{\rm BH} \sim 9$, with modest residuals. 

%However, the size scale defined by \citet{Mor10} typically underestimates the flux-weighted emission radius by a factor of 3 to 4 for visual light, and more for UV light from discs around very massive black holes (see Fig.~\ref{fig:t_M_lam}, left panel).

\begin{figure*}%[!htb]
\begin{center}
\includegraphics[width=0.85\columnwidth]{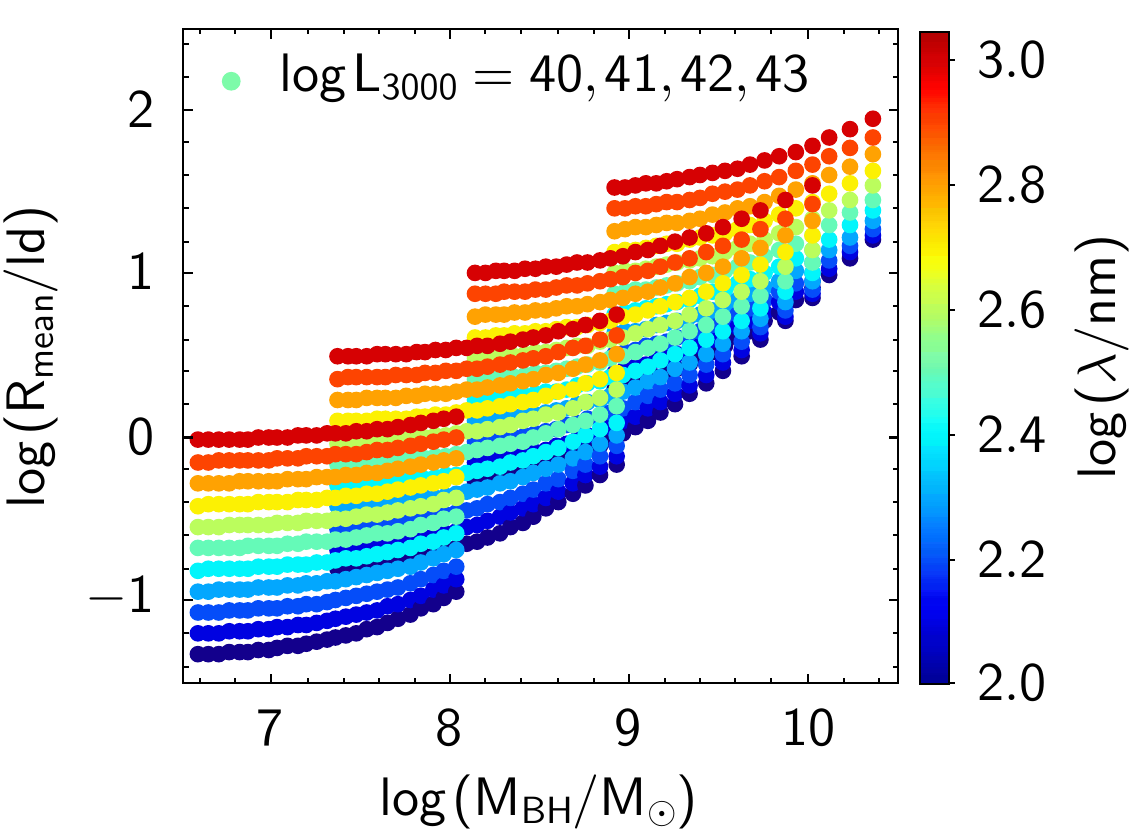}
\hspace{10mm}
\includegraphics[width=0.85\columnwidth]{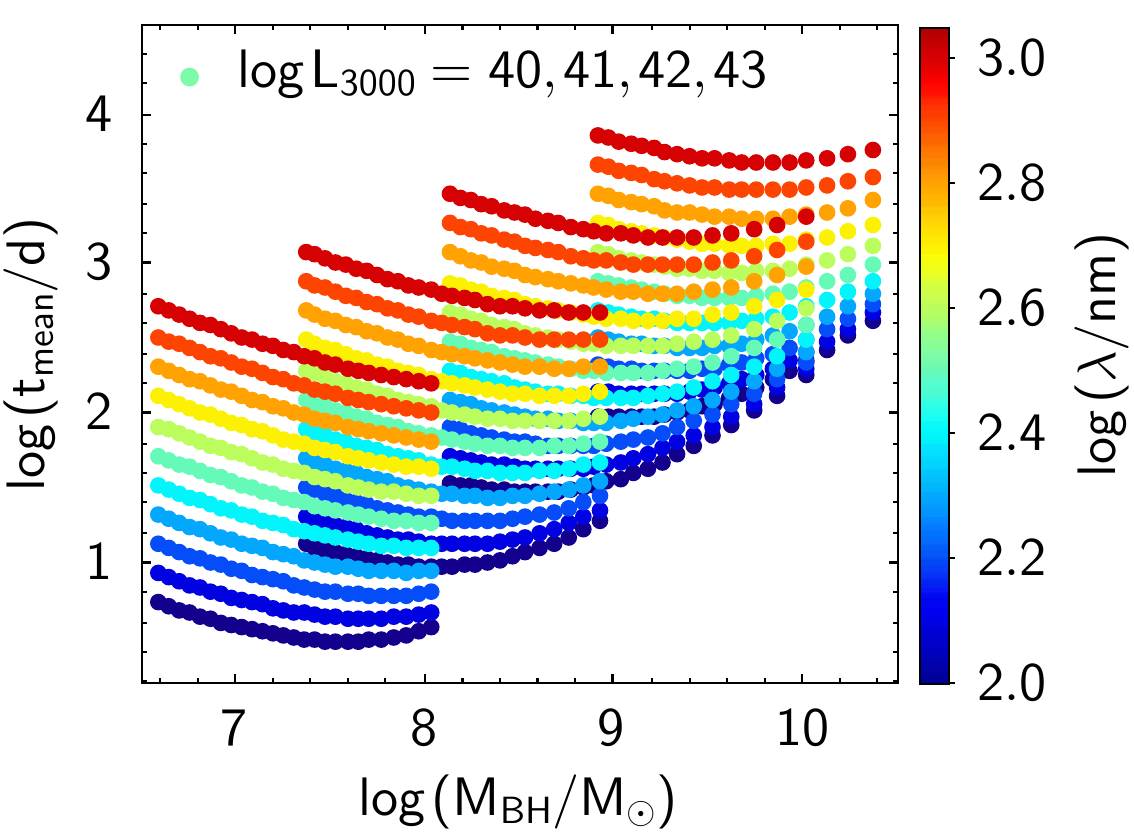}
\caption[]{Left: Flux-weighted size scale $R_{\rm mean}$ vs. black hole mass of four groups of discs with $\log L_{3000}/(\mathrm{erg\,s^{-1}}\,$\AA$^{-1})$ values of 40, 41, 42 and 43 ({\refbf from low $L_{3000}$ at low $M_{\rm BH}$ on the left to high $L_{3000}$ at high $M_{\rm BH}$ on the right}); the black hole masses range for each value of $L_{3000}$ is driven by the Eddington ratio limits of the calculated grid. Right: Flux-weighted orbital time scale $t_{\rm mean}$ vs. black hole mass of the same discs.
}\label{fig:t_M_lam}
\end{center}
\end{figure*}

\subsection{Scaling at fixed luminosity and wavelength dependence}

As stated, we consider the observable with the lowest uncertainty to be a monochromatic luminosity (ideally from a spectrum fit) such as $L_{3000}$, followed by black hole mass $M_{\rm BH}$ in second place. The Eddington ratio $R_{\rm Edd}$ comes last in this list, as it combines errors from the two previous observables and includes mass-dependent biases in the bolometric correction. Hence, we now consider the disc scaling behaviour with black hole mass at fixed observed $L_{3000}$.

{\refbf In Section~\ref{sec:intuition} we saw that at fixed $r_0$, all temperature profiles have the same shape except for scaling with $R_{\rm ref}\propto M_{\rm BH} \propto L^2_{3000}$. Conversely, at a fixed $L_{3000}$, a variation of $M_{\rm BH} \propto R_{\rm ISCO}$ varies the inner disc cutoff $r_0$ (see also  Figure~\ref{fig:Tprofile}).}
In Figure~\ref{fig:t_M_lam} we show again the mass dependence of the size and orbital time scales, but this time for just a few choices of observed luminosity $L_{3000}$ and instead several steps in wavelength. At low black hole mass, and thus small $R_{\rm ISCO}$, a change in mass has little effect on the extent and appearance of the disc and thus its size scale ({\refbf left} panel). But at intermediate masses, an increasing $R_{\rm ISCO} \propto M_{\rm BH}$ moves the inner disc edge outwards, pushing it against the small-mass $R_{\rm mean}$ and eventually driving $R_{\rm mean}$ out at a rate that will approximate $R_{\rm mean} \propto R_{\rm ISCO} \propto M_{\rm BH}$. 

For the orbital time scales (right panel), we then find at low masses, where size scales are nearly constant, that orbital velocity changes as $v \propto \sqrt{M_{\rm BH}}$ %^{1/2}
and thus time scales decline with increasing mass as $t_{\rm orb} \propto R/v \propto 1/\sqrt{M_{\rm BH}}$. At large masses, in contrast, the rapidly increasing size scale affects the orbital time scales more strongly than the declining orbital periods at fixed radius, causing $t_{\rm orb} \propto R/v \propto (R^3/M_{\rm BH})^{1/2} \propto M_{\rm BH}$. Between these two regimes, the orbital timescale reaches a minimum at a mass {\refbf $M_{\rm BH,tmin}$} that depends on luminosity and wavelength (see Figure~\ref{fig:t_M_lam}). At {\refbf$\log (\lambda /\textup{\AA})=3.5$} and $\log L_{\rm bol}/(\mathrm{erg}~\mathrm{s}^{-1})=47$, the minimum time scale is reached at $\log M_{\rm BH,tmin}\approx 9.5$. {\refbf This turnover means that the relation of orbital times scale to $M_{\rm BH}/\sqrt{L}$ is not unique, which is also the reason why $t_{\rm mean}/\sqrt{L}$ is not a unique second parameter. At fixed luminosity, the orbital time scale could be long because of the weak gravity of a low-mass black hole or because of a large flux-weighted orbital radius resulting from a large ISCO around a high-mass black hole.}

\begin{figure*}%[!htb]
\begin{center}
\includegraphics[width=\columnwidth]{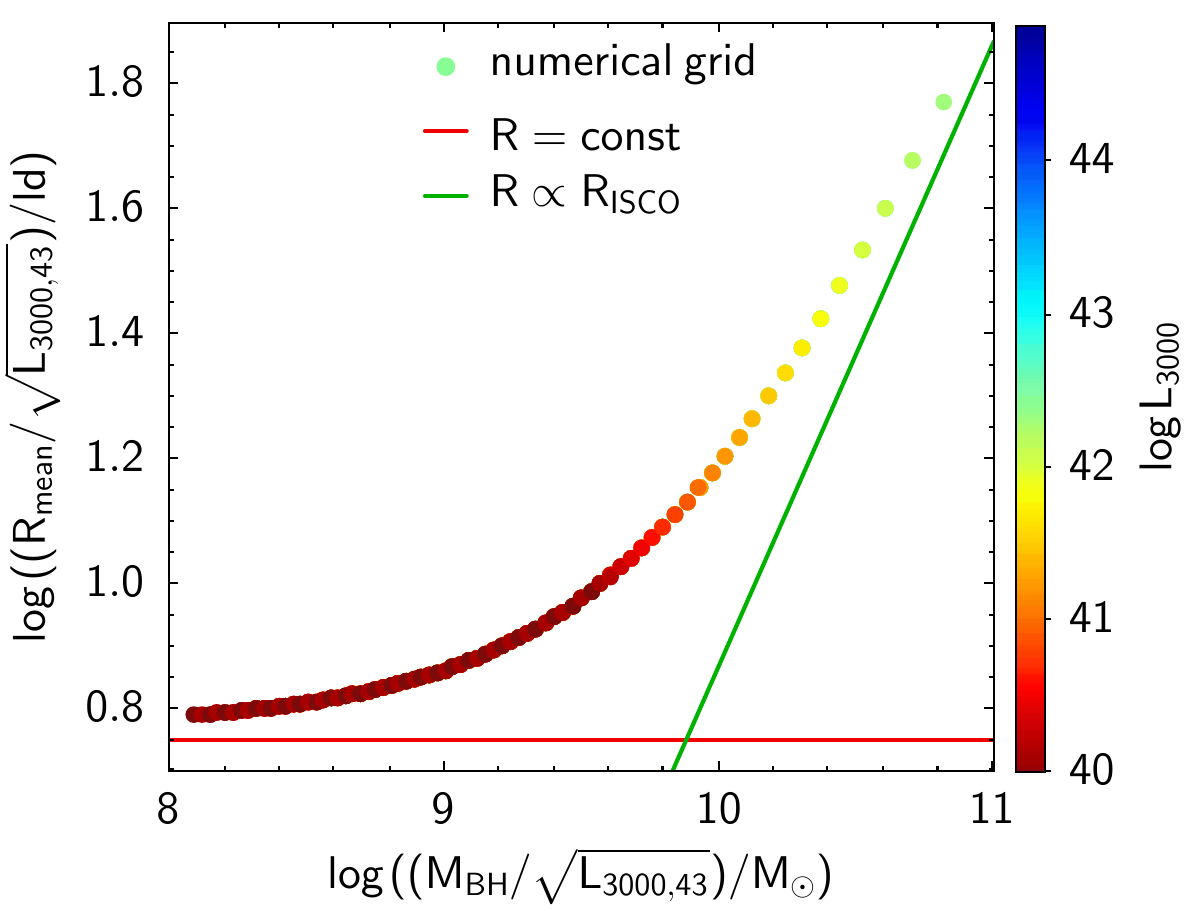}
\hspace{5mm}
\includegraphics[width=\columnwidth]{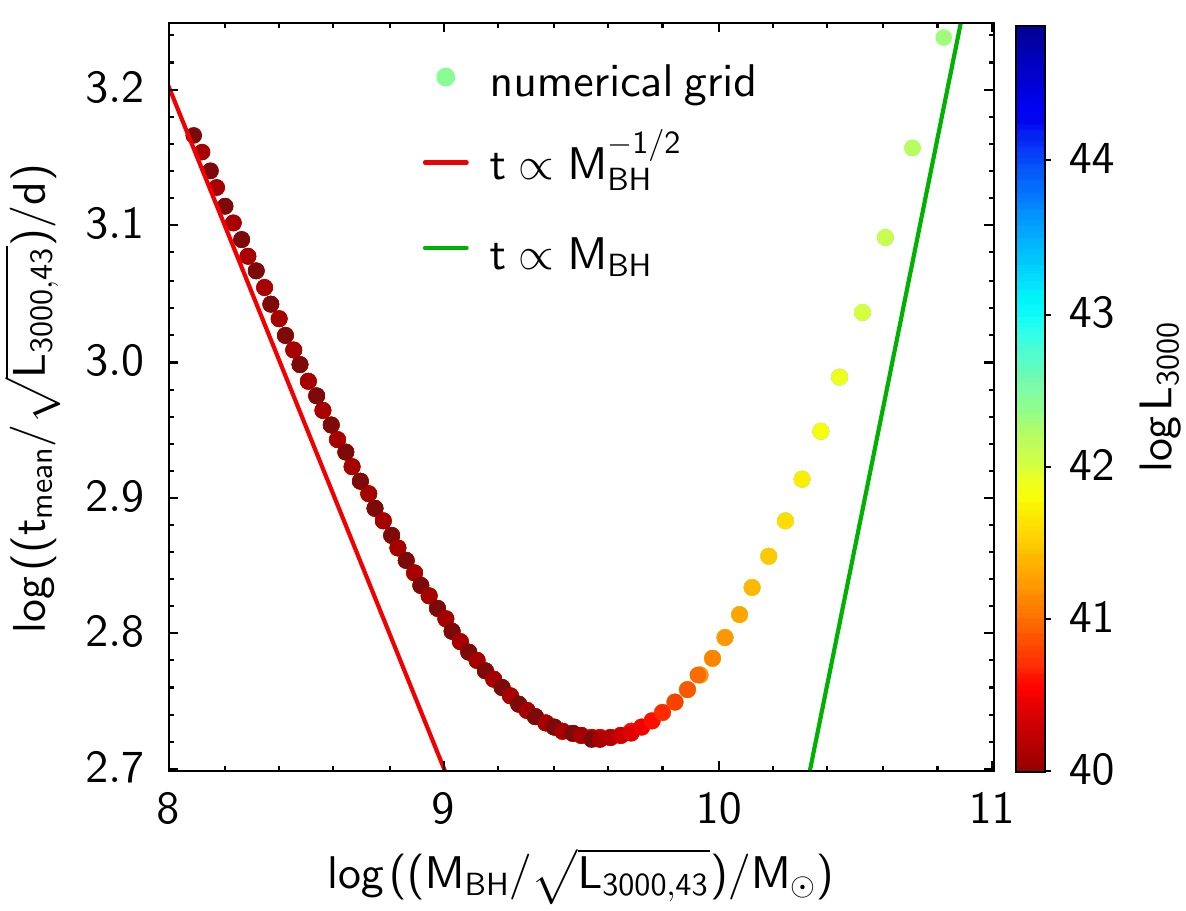} 
\caption[]{Luminosity-scaled size scales (left) and time scales (right) vs. luminosity-scaled black hole mass at emission wavelength {\refbf$\log (\lambda /\textup{\AA})=3.5$} for a range of accretion discs (GR approximation with spin $a=0$). For brevity, we use $L_{3000,43} = L_{3000}/(10^{43} \mathrm{erg\,s^{-1}\,}$\AA$^{-1})$. A warped 2D surface is seen in projection as a 1D line. Every point in this figure represents a 1-parameter family of discs with different $L_{3000}$ but identical $R_{\rm mean}/\sqrt{L_{3000}}$ and identical $t_{\rm mean}/\sqrt{L_{3000}}$. The variation of points seen in the projected plane is caused by variations in $R_{\rm ISCO}/R_{\rm mean}$. 
}\label{fig:rLtl_ML}
\end{center}
\end{figure*}

{\refbf
\subsection{Scaling with $M_{\rm BH}/\sqrt{L}$}

As we have established, at fixed black hole spin and fixed $r_0$ all discs have a fixed $R_{\rm ref}/\sqrt{L}$ and a fixed $M_{\rm BH}/\sqrt{L}$. In Figure~\ref{fig:t_M_lam}, a disc family with fixed $r_0$ will obviously appear at different locations $R_{\rm mean}(M_{\rm BH})$, but if instead we plot the invariant $y=R_{\rm mean}/\sqrt{L}$ as a function of the invariant $x=M_{\rm BH}/\sqrt{L}$, then the whole family will collapse into a single point. A 2D family of families with different $r_0$ will then appear as a 1D family in $y(x)$ as seen in Figure~\ref{fig:rLtl_ML},}
%We choose not to follow the parametrisation of \citet{Mor10} as 
%The simple model of a 1-parameter family scaled by size predicts a constant mean surface luminosity for a fixed temperature profile with inner and outer edges that scale in tune with the overall disc. In this case, we expect $R_{\rm mean}/ \sqrt{L}$ to be invariant, and at a fixed mass $t_{\rm mean}/\sqrt{L}$ as well. The outer disc edge can be defined as a fixed temperature point, which will scale in tune with the overall disc scale given the temperature point asymptotes to a common power law in temperature for all discs. The inner disc edge depends on black hole mass, so we can choose a pivotal mass as the one that keeps the inner edge scaling with the overall disc scale as well. Ignoring the effect of black hole spin for now, $R_{\rm ISCO} \propto M_{\rm BH}$, so that the pivotal mass could be chosen as $M_{\rm BH}/\sqrt{L}$. %However, with a free choice of mass we get a two-parameter family. 
%{\bf [the above is a bit waffled; ]}
%Thus, the 2-parameter family of discs spanned by an overall size scale and an inner-edge scale should appear as a 1-parameter family of $R_{\rm mean}/\sqrt{L}$ as a function of $M_{\rm BH}/\sqrt{L}$. This effect is confirmed by Figure~\ref{fig:rLtl_ML}, 
where $R_{\rm mean}/ \sqrt{L}$ and $t_{\rm mean}/\sqrt{L}$ are shown for discs of all {\refbf luminosities and black-hole masses. While they appear as a curved 2D surface in a $R_{\rm mean}(M_{\rm BH})$ diagram, they are seen in a 1D edge-on projection in the $R_{\rm mean}/\sqrt{L}$ {\it vs} $M_{\rm BH}/\sqrt{L}$ diagram. Note that} not all points are visible because the symbols are opaque and hiding points from the gridded surface in the background. 
At lowest and highest black hole masses the scales approach the analytically expected limiting behaviours. At low black hole mass, {\refbf varying a tiny ISCO makes a miniscule difference to the disc, and} %the spatial appearance of the discs is independent of the ISCO and the black hole mass,
the orbital time scale declines with $M_{\rm BH}^{-1/2}$; at high black hole mass, the appearance of the disc is driven by the {\refbf hole due to the} ISCO and thus black hole mass, such that the size scale will increase as $R_{\rm mean} \propto R_{\rm ISCO} \propto M_{\rm BH}$ and the orbital time scale with $t_{\rm mean} \propto R_{\rm ISCO}^{3/2} M_{\rm BH}^{-1/2} \propto M_{\rm BH}$. 

In the following section, we develop an analytic approximation to the numerically calculated surface by using a smoothly broken power law incorporating the outlined expected power law characteristics. {\refbf We note in anticipation of this, that at least in the low-mass regime, where the ISCO has little effect on the disc overall,} the dependence on wavelength should follow the temperature profile with roughly $R_{\rm mean} \propto \lambda^{4/3}$ and $t_{\rm mean} \propto R_{\rm mean}^{3/2} \propto \lambda^2$.

\section{A new approximation}\label{sec:approx}

We wish to assist future evaluations of size and time scales in simple thin-disc models by deriving an analytic approximation of $R_{\rm mean}=f(L_{3000},M_{\rm BH},\lambda)$ and $t_{\rm mean}=f(L_{3000},M_{\rm BH},\lambda)$ for different innermost stable orbits of $R_{\rm ISCO}/R_{\rm S}=(1.5,3,4.5)$ corresponding to black hole spins of $a=(+0.78,0,-1)$. These will not be single power laws but smoothly broken power laws that approximate the numerical calculations while morphing from typical low-mass scaling, where scales are independent of the ISCO, to typical high-mass scaling, where scales are driven by the ISCO. For brevity, we will use the notation $L_{3000,43} = L_{3000}/(10^{43} \mathrm{erg\,s^{-1}\,}$\AA$^{-1})$; {\refbf $\lambda$ will be in units of \AA \ and $M_{\rm BH}$ in units of $M_\odot$.} 
%{\bf CW: ADD t and R}

\iffalse
We use the general approach of smoothly broken power laws
\begin{equation}
   y = y_0 \times \left(\frac{x}{x_0}\right)^{s_1} \left[1+\left(\frac{x}{x_{\rm br}}\right)^{(s_2-s_1)/\gamma} \right]^\gamma ~,
\end{equation}
where $x=M_{\rm BH}/\sqrt{L_{3000}}$, the size scale is $y_r=R_{\rm mean}/\sqrt{L_{3000}}$ and the time scale is $y_t=t_{\rm mean}/\sqrt{L_{3000}}$ at fixed wavelength $\lambda_{\rm rest}$ and black hole spin $a$. For the size scale, $s_1=0$ and $s_2=1$, we get 
\begin{equation}
   y_r = C_r \times \left(\frac{\lambda}{\lambda_0}\right)^{4/3} \times \left[1+\left(\frac{x}{x_{\rm br}(\lambda)}\right)^{1/\gamma_r} \right]^{\gamma_r}  ~,
\end{equation}
and for the time scale, $s_1=-1/2$ and $s_2=1$, such that
\begin{equation}
   y_t = C_t \times \left(\frac{\lambda}{\lambda_0}\right)^2 \times \left(\frac{x}{x_0}\right)^{-1/2} \left[1+\left(\frac{x}{x_{\rm br}(\lambda)}\right)^{3/2\gamma_t} \right]^{\gamma_t}  ~.
\end{equation}

{\bf CW:} $\gamma(\lambda)$ is not obvious. Try $\gamma_t = 2/3 \gamma_r$. Could we try the below version to decouple the lambda dependence of PL-low from PL-high? Assuming we fix $\lambda_0=300$~nm to normalise, then we still have one more free parameter than if not: \\
\fi 

We use the general approach of smoothly broken power laws
\begin{equation*}   %remove * after settling matters
   y = y_0 \times \left[ \left( \left(\frac{x}{x_1}\right)^{s_1} \right)^\gamma + \left( \left(\frac{x}{x_2}\right)^{s_2} \right)^\gamma \right]^{1/\gamma} ~,
\end{equation*}
where $x=M_{\rm BH}/\sqrt{L_{3000,43}}$, the size scale is $y_r=R_{\rm mean}/\sqrt{L_{3000,43}}$ and the time scale is $y_t=t_{\rm mean}/\sqrt{L_{3000,43}}$ at fixed wavelength $\lambda_{\rm rest}$ and black hole spin $a$. After applying expected scaling behaviour and factoring in a reference wavelength $\lambda_0$, we get
\begin{equation}
   y_r = \left[ \left( C_r \left( \frac{\lambda}{\lambda_0}\right)^{4/3} \right)^{\gamma_r} + \left(\frac{x}{x_{\rm br,r}(\lambda)}\right)^{\gamma_r} \right]^{1/\gamma_r}  ~,\mathrm{ and}
\end{equation}
\begin{equation}
   y_t = \left[ \left( \left(\frac{\lambda}{\lambda_0}\right)^2 \left(\frac{x}{x_0}\right)^{-1/2}\right)^{\gamma_t} + \left(\frac{x}{x_{\rm br,t}(\lambda)}\right)^{\gamma_t} \right]^{1/\gamma_t}  ~.
\end{equation}

\begin{figure}%[!htb]
\begin{center}
\includegraphics[angle=0,width=0.95\columnwidth]{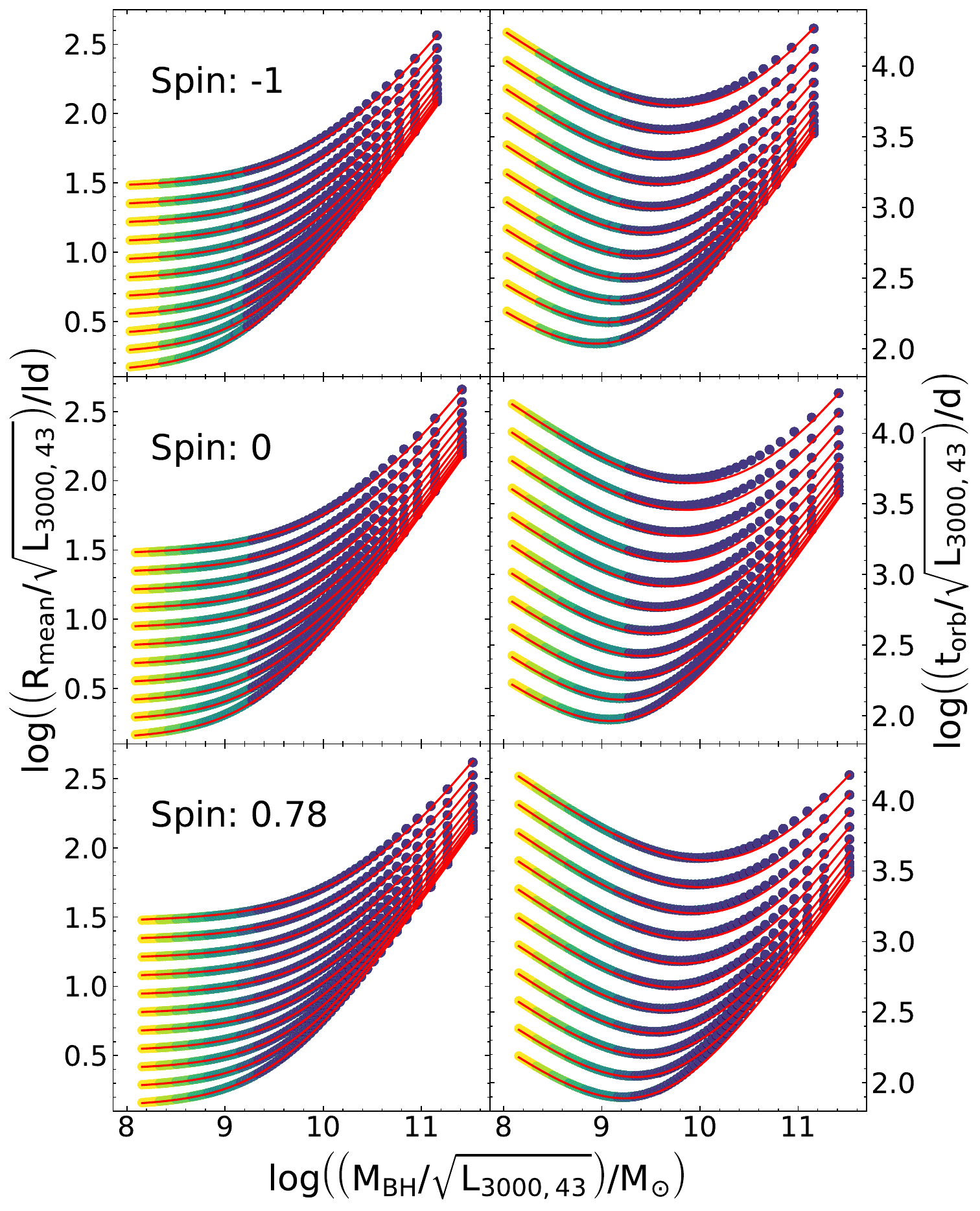}\\
\vspace{4mm}
\includegraphics[angle=0,width=0.95\columnwidth]{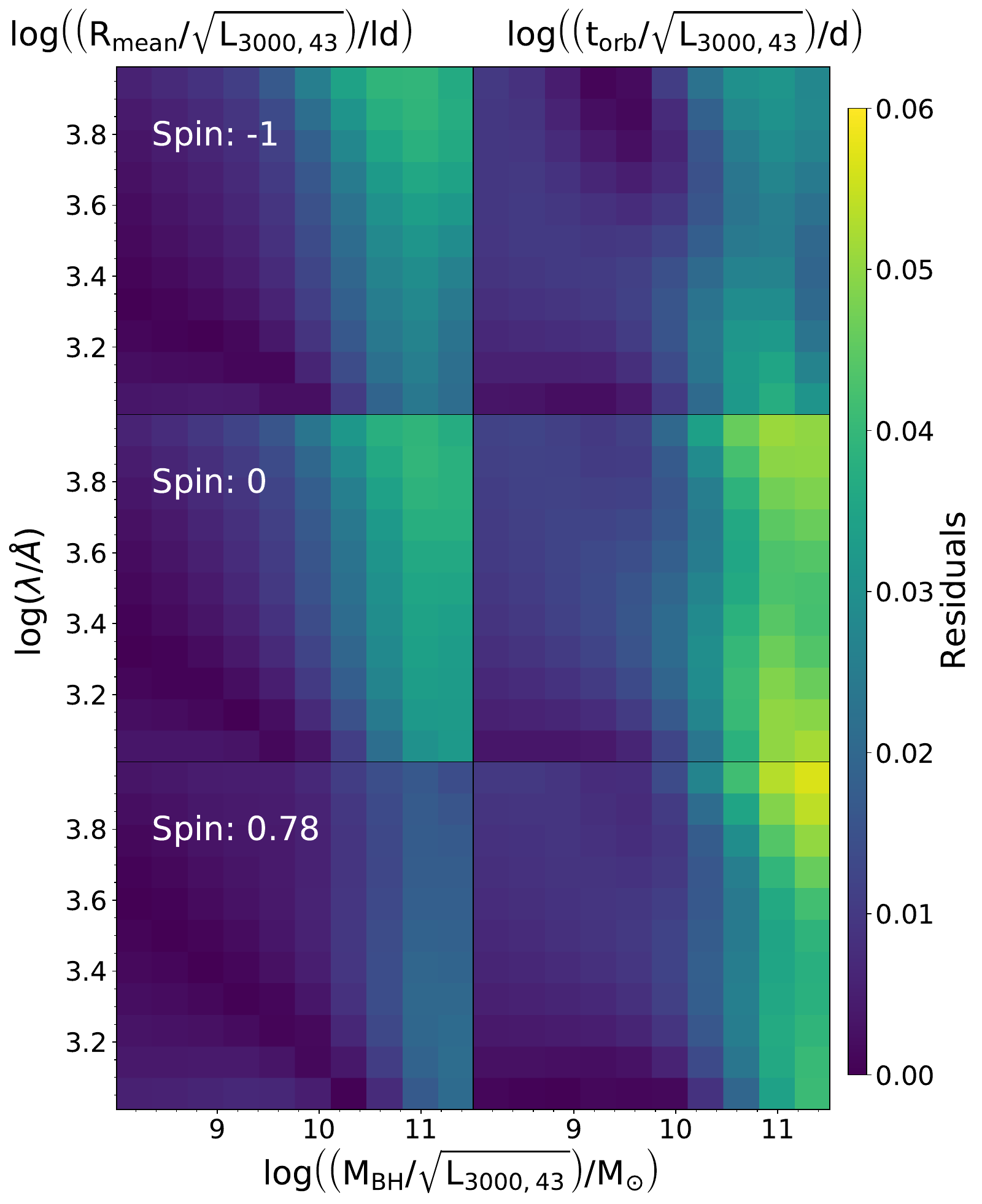}
\caption[]{Illustration of the analytic approximation to size and time scales in simple thin-disc models. {\it Top:} Grid points are displayed as points and the analytic fits are plotted as lines. The colour scale represents the range of Eddington ratios and lines in a single panel are differentiated by wavelength. {\it Bottom:} Residuals as a function of wavelength and mass, showing that while the analytic solution is typically within 0.01 dex of the numerical calculation, the deviation can reach 0.05 dex at high masses and low luminosities. 
}\label{fig:modelfit}
\end{center}
\end{figure}

After inspecting first results, we choose a further broken power-law parametrisation for the $\lambda$-dependence of $x_{\rm br,r}$ and proportionalities between size and time scale parameters. Our best-fit solution then is:
\begin{align}
    &\lambda_0 \equiv 3000\textup{\AA} \,, C_r \approx 5.896\,, x_0  \approx 2.279\times 10^{14}\,, \\
    &x_{\rm{br,r}}(\lambda, \hat a) =
     \left[\zeta(\hat a)^{\gamma_{\rm{br}}}+\left(\frac{\lambda}{\zeta(\hat a)\lambda_{\rm{br}}}\right)^{-\gamma_{\rm br}} \right]^{1/\gamma_{\rm br}}\,, \\
    &\log(\zeta(\hat a)) = 9.29 + 0.15\hat a\,,
    \log\lambda_{\rm{br}} = 3.66\,,
    \gamma_{\rm br} = -1.437\,, \\
%\end{align}
%\begin{align}
    &\log x_{\rm{br,t}} = \frac{3}{2}(\log x_{\rm{br,r}} - 4)\,,\\
    &\frac{3}{2}\gamma_{t}(\lambda) = \gamma_{r}(\lambda) = C_0 + C_1\log\lambda\,,\\
    &C_0 =  1.683\,, C_1 = -0.246\,, \hat a = (6-2R_{\rm ISCO}/R_{\rm S})/3 ~.
\end{align}
% $r_{\rm ISCO}= R_{\rm ISCO}/R_{\rm S}$
%{\bf FIX the above, in units of $(6-2r_{\rm ISCO})/3$ instead of $a$}

This parametric solution agrees with the numerical calculations to $<0.01$~dex for most of the grid range in both $\log R_{\rm mean}$ and $\log t_{\rm mean}$; however, at large masses and small luminosities, the deviation can reach 0.05~dex; Figure~\ref{fig:modelfit} shows the quality and residuals of the fit.

\section{Conclusions}\label{sec:conc}

{\refbf Many literature studies are concerned with how the scaling behaviour of stochastic UV-optical variability in AGN depends on AGN parameters. Some works consider a dependence on black hole mass \citep[including most recently][]{Arevalo24}, while others ignore this dependence \citep[including ][]{TWT23}.} A specific dependence considered since \citet{BH91} and \citet{Kelly09} is that variability behaviour may depend on orbital or thermal timescales of the emitting accretion disc. %The question then is how these depend on black hole mass and whether that matches observations. 
{\refbf Their dependence on black-hole mass is often approximated by a power law, but in this paper we reveal that emission-weighted size and time scales depend on black-hole mass in a non-trivial way best represented by two regimes of behaviour. % with a smooth transition region in between
}

We first model standard thin accretion discs and evaluate mean orbital timescales of the disc over the following parameter ranges: the rest-frame wavelength of disc emission, $\log (\lambda_{\rm rest}/$\AA $)=[3;4]$, the black hole mass, $\log (M_{\rm BH}/M_\odot)=[6;11]$, the Eddington ratio of the disc, $\log (R_{\rm Edd})=[-2;0]$, and black hole spin values of $a=(+0.78,0,-1)$. Before studying dependencies, we calculate the monochromatic 3\,000\AA \ disc luminosity, $L_{3000}$, which is a more robustly determined observable than $R_{\rm Edd}$.

{\refbf Our calculations show that the quantity $x=M_{\rm BH}/\sqrt{L_{3000}}$ is a practical ordering parameter for accretion discs around supermassive black holes, given that the size and time scales of $y_r=R_{\rm mean}/\sqrt{L_{3000}}$ and $y_t=t_{\rm mean}/\sqrt{L_{3000}}$ are fixed for a given value of $x$. Accretion discs with different luminosities $L_{3000}$ are self-similar as long as they are paired with black holes of mass $M_{\rm BH}\propto \sqrt{L_{3000}}$. While varying $x$}, we find two regimes in the timescale dependence on black hole mass, with a turnover in between: at low masses, we see the decline of $t_{\rm orb} \propto M^{-1/2}$, which is a textbook expectation of orbits speeding up with increasing central mass. Towards extremely massive black holes, we observe that a growing event horizon and innermost stable circular orbit (ISCO) around the black hole push the emission region farther from the black hole such that we see an increase in timescale with mass, $t_{\rm orb} \propto M$. These two regimes are connected by a transition region, where the mass dependence vanishes locally. The relation between disc timescale and black hole mass is thus not a simple power law but a smoothly broken power law. For the benefit of the reader, we approximate the numerical grid model with convenience functions that express the mean emission radius and the mean orbital timescale as a function of wavelength, black hole mass, monochromatic luminosity, and black hole spin. 

It might come as a surprise that observed quasar samples {\refbf reach the transition regime and perhaps the rising branch of the timescales}. The black hole mass that minimises disc time scales for a disc with $\log L_{\rm bol}/(\mathrm{erg}~\mathrm{s}^{-1})=47$ is $\log M_{\rm BH}\approx 9.5$. {\refbf \citet{TWT23} and \citet{Tang24b} targeted samples of a few thousand of the most luminous known quasars, with median $\log L_{\rm bol}/(\mathrm{erg}~\mathrm{s}^{-1})\approx 47$ and a median black-hole mass of $\log M_{\rm BH}\approx 9.3$. If the black-hole mass estimates for these quasars are not systematically and strongly overestimated, then we expect their emission-weighted time scales} to exhibit little mass dependence. In hindsight, this may justify that \citet{TWT23} chose to ignore a black hole mass dependence in their estimates of disc timescales. {\refbf A caveat is that the luminosities might be underestimated if moderately extinguished by dust, and black-hole masses might be overestimated in the high-luminosity extrapolations of common mass estimators.}

{\refbf The convenience functions for disc time scales presented here will assist future studies of quasar variability with relating observed characteristic timescales to estimates of physical disc timescales. Persistent limits in our understanding of the structure of accretion discs in quasar and the physics of their variability ensure strong continued interest in higher}-precision observations of UV-optical variability in quasar discs {\refbf regardless of specific theories for their interpretation. Such observations will be carried out by} the Legacy Survey of Space and Time \citep[LSST;][]{Ivezic2008} at the Vera C.~Rubin Telescope in Chile. For brighter quasars that saturate in LSST observations, NASA/ATLAS \citep{To18a}, the Zwicky Transient Facility \citep[ZTF;][]{Bellm19} and others will continue to play a role, {\refbf although their range of spectral passbands is limited. In a follow-up paper, we will analyse structure functions of an enlarged sample of quasars with updated longer light curves from ATLAS.
}

\section*{Acknowledgements}

JJT was supported by the Taiwan Australian National University PhD scholarship 
%and the Australian Research Council (ARC) through Discovery Project DP190100252 
and by the Institute of Astronomy and Astrophysics, Academia Sinica (ASIAA). This research has made use of \textsc{idl}. We thank Zachary Steyn for helping to remove ambiguities in the manuscript.
\section*{Data Availability}
The data underlying this article will be shared on reasonable request to the corresponding author. 
%The inclusion of a Data Availability Statement is a requirement for articles published in MNRAS. Data Availability Statements provide a standardised format for readers to understand the availability of data underlying the research results described in the article. The statement may refer to original data generated in the course of the study or to third-party data analysed in the article. The statement should describe and provide means of access, where possible, by linking to the data or providing the required accession numbers for the relevant databases or DOIs.

\section*{Appendix}

\begin{figure*}%[!htb]
\begin{center}
\includegraphics[width=\textwidth]{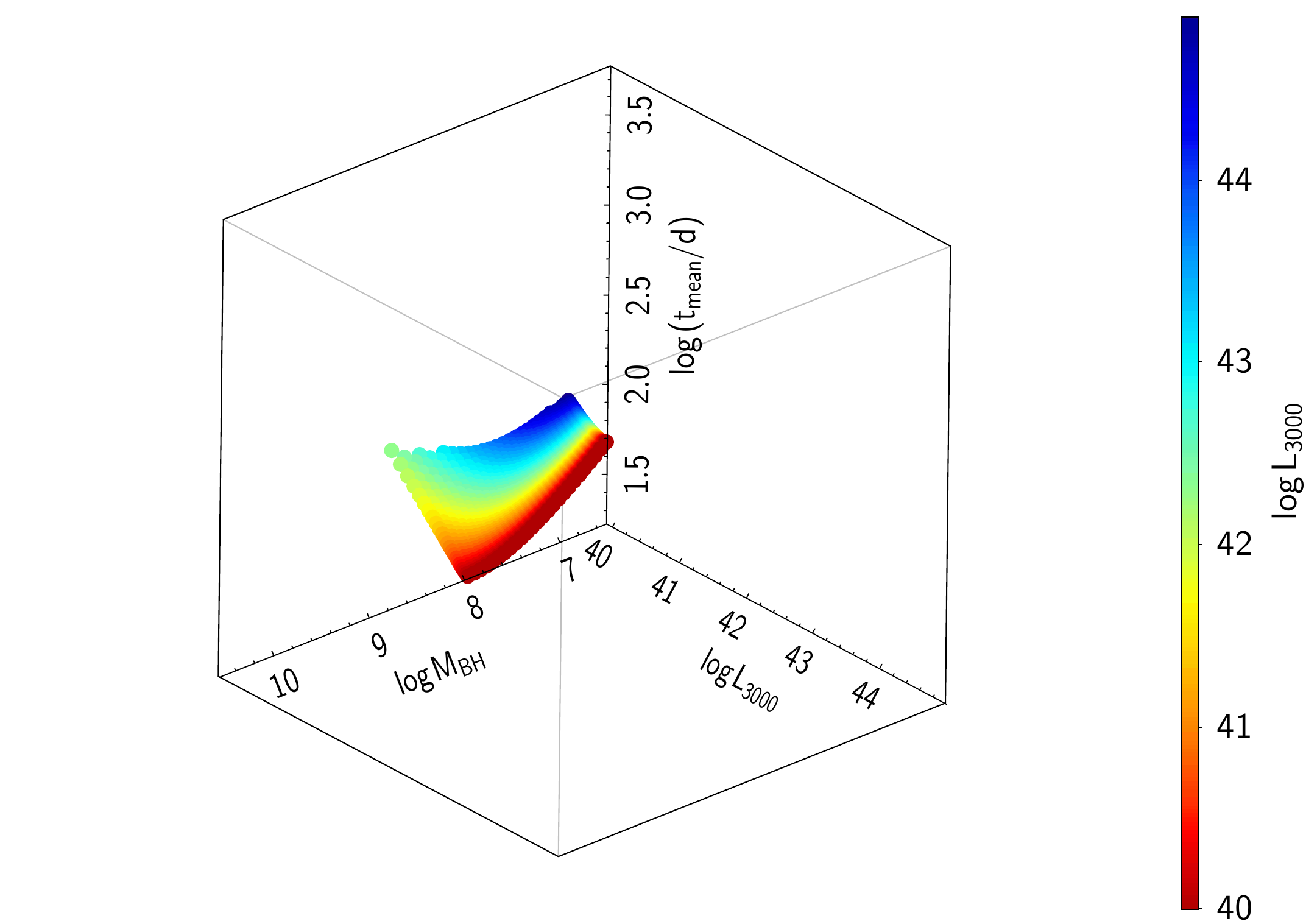}
\caption[]{A 3D view of the time scale vs. luminosity and black hole mass, for a wavelength of {\refbf$\log (\lambda /\textup{\AA})=3.5$}, illustrating the curvature of the time-scale plane. Figure~\ref{fig:rLtl_ML}, right panel, offers a view of this curved plane from a diagonal perspective that renders it as a 1D curve seen edge-on.
}\label{fig:cube3d}
\end{center}
\end{figure*}

Figure~\ref{fig:cube3d} shows a 3D view of the disc time scale in the luminosity-mass plane as calculated in the numerical grid. In the right panel of Figure~\ref{fig:rLtl_ML}, this plane is shown in a coordinate system rotated such that the curved plane is seen edge-on and appears as a 1D curve.

%Notes for working with Ji-Jia's cubes: \\

%\texttt{Lbol = 38.097+MBH+EDD\_RAT} \\
%\texttt{EDD\_RAT\_est = L3000-42.908+(MBH-9)} \\
%\texttt{RM10 = 0.5734+4./3.*(WV-4)+2./3.*(MBH-9)+\\ 1./3.*(EDD\_RAT-log10(0.057))} \\

%%%%%%%%%%%%%%%%%%%% REFERENCES %%%%%%%%%%%%%%%%%%

% The best way to enter references is to use BibTeX:

\bibliographystyle{mnras}
\bibliography{size}

% Alternatively you could enter them by hand, like this:
% This method is tedious and prone to error if you have lots of references
%\begin{thebibliography}{99}
%\bibitem[\protect\citeauthoryear{Author}{2012}]{Author2012}
%Author A.~N., 2013, Journal of Improbable Astronomy, 1, 1
%\bibitem[\protect\citeauthoryear{Others}{2013}]{Others2013}
%Others S., 2012, Journal of Interesting Stuff, 17, 198
%\end{thebibliography}

%%%%%%%%%%%%%%%%%%%%%%%%%%%%%%%%%%%%%%%%%%%%%%%%%%

%%%%%%%%%%%%%%%%% APPENDICES %%%%%%%%%%%%%%%%%%%%%

%\onecolumn

\begin{appendix}

\end{appendix}

%%%%%%%%%%%%%%%%%%%%%%%%%%%%%%%%%%%%%%%%%%%%%%%%%%

% Don't change these lines
\bsp	% typesetting comment
\label{lastpage}
\end{document}